\documentclass[aps,prd,onecolumn,amsmath,showpacs,showkeys,superscriptaddress,nofootinbib,nopreprintnumbers,tightenlines,notitlepage]{revtex4-1}

\pdfoutput=1

\usepackage{verbatim}
\usepackage[T1]{fontenc}
\usepackage[utf8]{inputenc}
\usepackage[american]{babel}
\usepackage{epsfig}
\usepackage{graphicx}

\usepackage{hyperref}

\usepackage{booktabs}
\usepackage{multirow}
\usepackage{dcolumn}
\usepackage{amsmath}
\usepackage{mathtools}
\usepackage{amsfonts}
\usepackage{amssymb}
\usepackage{epstopdf}
\usepackage{bm}
\usepackage{siunitx}
\usepackage{braket}
\usepackage{enumitem}
\usepackage{soul}
\usepackage[table]{xcolor}
\usepackage{color}
\usepackage{transparent}

\usepackage{enumitem}
\usepackage{graphicx}
\usepackage[font=small]{caption}
\usepackage{subcaption}
\usepackage{pifont}

\usepackage{todonotes}
\providecommand{\U}[1]{\protect\rule{.1in}{.1in}}

\newcommand{\beq}{\begin{eqnarray}}
\newcommand{\eeq}{\end{eqnarray}}
\newcommand{\bea}{\begin{eqnarray*}}
\newcommand{\eea}{\end{eqnarray*}}
\newcommand{\eq}{eqnarray}

\newcommand{\al}{{\alpha}}
\newcommand{\be}{{\beta}}
\newcommand{\ci}{\cite}
\newcommand{\ga}{{\gamma}}

\newcommand{\ep}{{\epsilon}}

\newcommand{\de}{{\delta}}
\newcommand{\De}{\Delta}

\newcommand{\la}{{\lambda}}
\newcommand{\La}{{\Lambda}}
\newcommand{\m}{{\mu}}

\newcommand{\s}{{\sigma}}
\newcommand{\Si}{{\Sigma}}
\newcommand{\ka}{{\kappa}}
\newcommand{\om}{{\omega}}
\newcommand{\Om}{{\Omega}}

\newcommand{\no}{{\nonumber}}
\newcommand{\f}{\frac}
\newcommand{\ra}{\rightarrow}
\newcommand{\lra}{\leftrightarrow}

\newcommand{\Ho}{Ho\v{r}ava}

\begin{document}
\title{A New Test of Dynamical Dark Energy Models
and Cosmic Tensions in Ho\v{r}ava Gravity}

\author{Eleonora Di Valentino}
\email{e.divalentino@sheffield.ac.uk}
\affiliation{School of Mathematics and Statistics, University of Sheffield, Hounsfield Road, Sheffield S3 7RH, United Kingdom}

\author{Nils A. Nilsson}
\email{nilssonnilsalbin@gmail.com}
\affiliation{SYRTE, Observatoire de Paris-PSL, Sorbonne Universit\'e, CNRS UMR8630, LNE, 61 avenue de
l’Observatoire, 75014 Paris, France}
\affiliation{Center for Quantum Spacetime (CQUeST), Sogang University, Seoul, 121-742, Korea}

\author{Mu-In Park}
\email{muinpark@gmail.com, Corresponding author}
\affiliation{Center for Quantum Spacetime (CQUeST), Sogang University, Seoul, 121-742, Korea}

\begin{abstract}
{Ho\v{r}ava gravity has been proposed as a renormalizable, higher-derivative, Lorentz-violating quantum gravity model without ghost problems.
A Ho\v{r}ava gravity based dark energy ({\it HDE}) model for dynamical dark energy has 
been also proposed earlier by identifying all the extra (gravitational) 
contributions from the Lorentz-violating terms as an {\it effective} 
energy-momentum tensor in Einstein equation. We consider a complete CMB, BAO, and SNe Ia data test of 
the HDE model by considering general perturbations over the background 
perfect HDE fluid. Except from BAO, we obtain the preference 
of {\it non-flat} universes for all other data-set combinations.
We obtain a positive result on the cosmic tensions between
the Hubble constant $H_0$ and the cosmic shear $S_8$, because we have a shift of $H_0$ towards a higher value, though not enough for resolving the $H_0$ tension, but the value of $S_8$ is unaltered. This is in contrast to a rather decreasing $H_0$ but increasing $S_8$ in a non-flat LCDM.
For all other parameters, like $\Omega_m$ and $\Omega_\Lambda$, we obtain quite comparable results with those of LCDM for all data sets, especially with BAO, so that our results are close to a {\it cosmic concordance} between the datasets, contrary to the standard non-flat LCDM.
We also obtain some undesirable features{, like an almost {\it null} result on $\Om_k$, which gives back the flat LCDM, if we do not predetermine the sign of $\Om_k$,} but we propose several promising ways for improvements by generalizing our analysis.}
\end{abstract} 


\maketitle

\section{Introduction}
\label{sec:intro}

In 2009, Ho\v{r}ava proposed a renormalizable,
higher-derivative, Lorentz-violating quantum gravity model
without the ghost problem, due to anisotropic scaling dimensions for space and time
{\it \`{a} la} Lifshitz and DeWitt~\cite{Lifs:1941,DeWi:1967,Hora:2009}. In the last 12 years, there have
been a lot of work on its various aspects (see \ci{Wang:2017} for a brief review and extensive literature). In particular, in~\cite{Park:2009}, the author interpreted the {\it dark energy} as an {\it effective} energy-momentum in Einstein equation due to
the extra contributions from the Lorentz-violating terms.

The Ho\v{r}ava gravity based dark energy ({\it HDE}) model explains naturally the non-interacting nature of the dark energy sector, except the gravitational interactions since it was originally a part of the gravity sector, with the ordinary matter sector. Furthermore, it also predicts {\it dynamical} dark energy behavior in the cosmic evolution, depending on the purported Ho\v{r}ava  gravity action, which may contain various spatially-higher-derivative (UV) Lorentz-violating terms (up to sixth order or $z=3$ in $(3+1)$ dimensions for renormalizability \ci{Hora:2009}) and IR Lorentz-violating terms.
A peculiar property of HDE is that a spatially non-flat universe may be more
``natural'' due to genuine contributions from higher-spatial derivatives~\cite{Park:2009} since, for a spatially flat universe, the usual Friedmann-Lemaitre-Robertson-Walker (FLRW) {\it background} cosmology~\cite{Frie:1922,Lema:1927} is the same as in GR. In other words, Ho\v{r}ava gravity can be a ``natural laboratory'' for the test of a non-flat universe in the standard Lambda Cold Dark Matter (LCDM) cosmology.

On the other hand, with the increased precision of cosmological data, some
cosmic tensions between different data sets are becoming clearer~\cite{Abdalla:2022yfr,DiValentino:2020zio,DiValentino:2020vvd,Shah:2021onj} within the standard LCDM paradigm and its complete resolution is a challenging problem in current cosmology. In particular, there have been various proposals~\cite{Knox:2019rjx,Jedamzik:2020zmd,DiValentino:2021izs,Perivolaropoulos:2021jda,Kamionkowski:2022pkx} trying to address the tensions involving the Hubble constant
$H_0$~\cite{Riess:2021jrx,Verde:2019ivm}
and $S_8\equiv \s_8 \sqrt{\Om_m/0.3}$, measured by cosmic shear experiments~\cite{KiDS:2020suj,DES:2021wwk},
between Cosmic Microwave Background (CMB) data and local measurements at lower redshifts, which corresponds to the
{\it mismatches between the early and late universe},
but a resolution at the fundamental level is still missing.
Moreover, when the possibility of a {\it closed} universe is considered~\cite{DiVa:2019,Hand:2019,DiValentino:2020srs,DiValentino:2020hov,Yang:2022kho,Semenaite:2022unt}, the tensions become worse, due to a decreasing $H_0$ but increasing $S_8$, and a new problem appears, called the {\it discordance} problem which means `the lack of {\it concordance} -- in other words, the lack of consistency -- with other observations'; for example, the closed universe preferred by Planck predicts $\Om_m \sim 0.5$ and consequently $\Om_\La \sim 0.5$, in sharp contrast to the conventional value $\Om_m \sim 0.3$ and $\Om_\La \sim 0.7$ for the local measurements or flat LCDM.

This motivated the authors
of~\cite{Nils:2021} to analyze current cosmological data, including Baryon Acoustic Oscillations (BAO), for the
standard (background) FLRW cosmology with the
HDE model and found some improvements of the Hubble constant tension with a preference
of a ``closed'' universe but {\it without} the problems of too large $\Om_m$ and too small $\Om_\La$ in a non-flat LCDM, {\it i.e.}, improving the discordance problem. However, the previous
analysis was not complete in two aspects:
{\it First},
due to the lack of {\it perturbations} of matter and dark energy, one can not spell out the $S_8$ tension from the amplitude $\s_8$ of matter density fluctuations on scales of 8 Mpc $h^{-1}$;
{\it Second}, due to the {\it algorithmic} limitation of the Metropolis-Hastings algorithm~\cite{Robe:2015}, the convergent $\chi^2$ statistics for separate data sets was not possible. From this second limitation, $\Om_m$ and $\Om_\La$ were obtained only for {\it all} the data sets whose reasonable value seems to support the lack of the discordance problem for {\it separate} data sets, but its explicit confirmation in each data set is still absent.

In this paper, in order to fill the gap, we extend the background analysis by considering {\it dark-energy perturbations} and using the full CMB data via CAMB/CosmoMC.
In Sec. II, we consider the theoretical setup for the cosmological perturbations of the Ho\v{r}ava gravity based dark energy (HDE) model on a {\it non-flat} FLRW background, based on the fluid approach for the perturbed HDE. We compare our HDE model whose equation of state (EoS) parameter is rapidly varying or fluctuating with the standard CPL model and show a good agreement in the comoving angular diameter distance, which supports the robustness of our HDE model in analyzing observational data. We also present the implicit assumptions in our fluid approach and initial conditions for solving perturbation equations. In Sec. III, we describe our methodology for the analysis and the data sets under consideration. In Sec. IV, we present and discuss our obtained results. In Sec. V, we conclude by proposing several promising directions to improve our analysis.

\section{Theoretical Setup}

\subsection{Dynamical Dark Energy Model in Ho\v{r}ava Gravity: HDE model}

We start by briefly reviewing the dynamical dark energy model in Ho\v{r}ava gravity, named HDE model~\cite{Park:2009}. To this ends, we consider the ADM (Arnowitt-Deser-Misner~\cite{Arnowitt:1962hi}) metric
\begin{\eq}
ds^2=-N^2 c^2 d t^2
+g_{ij}\left(dx^i+N^i dt\right)\left(dx^j+N^j dt\right)\,
\end{\eq}
and the ({non-projectable} \footnote{{
In the {\it projectable} case~\cite{Hora:2009,Muko:2009}, where the lapse function $N$ is a function of time only, there exists one extra scalar graviton mode. But in this paper,
we will not consider those cases in order to recover GR at the low-energy (or IR) limit, not to mention its pathological ghost behavior 
\cite{Bogdanos:2009uj,Koyama:2009hc,Cerioni:2010uz} which is though somewhat improved in the extended model
with the dynamical lapse function~\cite{Blas:2009}. Even in the {\it non-projectable} case, where $N$ is a function of space as well as time generally, it has been also known to have similar problems~\cite{Blas:2009yd} which have been the motivation for the extended model in~\cite{Blas:2009}, but it has been later found that there is no extra graviton mode problem in cosmological perturbations at the linear order~\cite{Gao:2009ht,Shin:2017}.} }) Ho\v{r}ava gravity action {\it \`{a} la} Lifshitz, DeWitt (HLD)~\cite{Lifs:1941,DeWi:1967,Hora:2009,Shin:2017},
given by
\begin{\eq}
\label{HL action}
S_\mathrm{HLD} &=& \int d t d^3 x \sqrt{g} N \left[ \frac{2}{\kappa^2}
\left(K_{ij}K^{ij} - \lambda K^2 \right) - {\cal V} \right], \\
-{\cal V}&=& \sigma+ \xi R + \alpha_1 R^2+ \alpha_2 R_{ij}R^{ij}
+\alpha_3 \frac{\epsilon^{ijk}}{\sqrt{g}}R_{il}\nabla_jR^l{}_k \no \\
 &+& \alpha_4 \nabla_{i}R_{jk} \nabla^{i}{R}^{jk}
+\al_5 \nabla_{i}R_{jk}\nabla^{j} {R}^{ik}
+\al_6 \nabla_{i}R\nabla^{i}R , 
\end{\eq}
which is power-counting renormalizable without the ghost problem in physical {\it TT} (transverse-traceless) graviton modes. Here,
\begin{\eq}
K_{ij}=\frac{1}{2N}\left(\dot{g}_{ij}
-\nabla_i N_j-\nabla_jN_i\right)\
\end{\eq}
is the extrinsic curvature (the dot $(\dot{~})$ denotes the
time derivative with respect to physical time $t$), $\nabla_i$ is the covariant
derivative with respect to $3$-metric $g_{ij}$, $R_{ij}$ is
the Ricci tensor of a
three-geometry, $K=g_{ij} K^{ij}, R=g_{ij} R^{ij}$ are their traces,
$\ep^{ijk}$ is the Levi-Civita symbol, and $\kappa,\lambda,\xi,\alpha_i,\s$ are coupling constants.

From the ``detailed balanced condition'' (DBC), which was adopted from quantum critical phenomena in condensed matter systems~\cite{Hora:2009}, the number of independent coupling constants reduce to six, {\it i.e.}, $\ka, \la, \mu, \nu, \La_W, \om$ for a viable gravity theory in the IR~\cite{Keha:2009,Park:2009} and the theory parameters are given by
\begin{\eq}
\label{parameters}
&&\sigma=\f{3 \kappa^2 \mu^2 \La_W^2}{8 (3 \la-1)},~
\xi=\f{\ka^2 \mu^2 (\om-\La_W)}{8 (3 \la-1)},~
\al_1=\f{\ka^2 \mu^2 (4 \la-1)}{32 (3 \la-1)},~
\al_2=-\f{\ka^2 \mu^2 }{8} \no \\
&&\al_3=\f{\ka^2 \mu }{2 \nu^2},~
\al_4=-\f{\ka^2}{8 2 \nu^4}=-\al_5=-8 \al_6,
\end{\eq}
with $\mu^2>0 ~(<0)$ for a positive (negative) cosmological constant $(\sim \La_W)$. Here, DeWitt's IR Lorentz-violation parameter $\la$~\cite{DeWi:1967} can be arbitrary, but below we will restrict to the case $\la=1$ as in GR 
{so that our analysis at the low energy agrees with the standard analysis in GR.} 
The UV couplings $\al_3,\cdots, \al_6$ do not appear explicitly in our analysis below but we will discuss their interesting role in the dark energy perturbations in Sec. V.

Then we may consider the gravity equations of motion for our universe with matter (ordinary and dark matter, and radiation) as
\begin{\eq}
G^{\mu \nu}=\f{8 \pi G}{c^4} \left(T^{\mu \nu}_{\rm matter}+ T^{\mu \nu}_{\rm DE} \right)
\label{Einstein_eq}
\end{\eq}
for the Einstein tensor $G^{\mu \nu}=\hat{R}^{\mu \nu}-(1/2)\hat{g}^{\mu \nu} \hat{R}$ with the $(3+1)$-dimensional Ricci tensor $\hat{R}^{\mu \nu}$, Ricci scalar $\hat{R}$, and covariant derivative $\hat{\nabla}_{\mu}$ by treating all the Lorentz-violating contributions from the HLD action (\ref{HL action}) as an ``effective'' dark energy {\it fluid} with the energy-momentum tensor $T^{\mu \nu}_{\rm DE}$, including the cosmological constant term~\cite{Park:2009}.
This interpretation is based on the fact that {the} dark energy is defined by the ``unknown'' contributions, other than dark matter, in our universe when described by GR. An important direct consequence of the interpretation is the (usually assumed) {\it non-interacting} nature of dark energy, except the gravitational interactions, with matter can be easily explained.\footnote{If there is a way to identify the dark matter as well as the dark energy from the gravity sector, one can also explain the non-interaction of dark matter with ordinary matter and radiation but with a possible interaction of dark energy and dark matter~\cite{Gave:2009}. In { projectable} Ho\v{r}ava gravity, cold dark matter (CDM) can be naturally introduced as an ``integration constant''~\cite{Muko:2009}. If one can realize the similar CDM behavior in its non-projectable version with  ``$a_i$-extended terms''~\cite{Blas:2009}, which is related to {\it Einstein Aether}~\cite{Jaco:2004} or {\it Standard-Model-Extension} (SME) gravity at low energy~\cite{ONea:2020}, one can explain the null result in direct detection of dark matter via particle interactions (see Sec. V, No. 2 for further discussions).} Moreover, from the Bianchi identity $\hat{\nabla}_{\mu} G^{\mu \nu}=0$ and the covariant conservation law of matter $\hat{\nabla}_{\mu}
T^{\mu \nu}_{\rm matter}=0$, one can find the conservation of dark energy $\hat{\nabla}_{\mu} T^{\mu \nu}_{\rm DE}=0$ as well, even with the Lorentz-violating terms of the HLD action (\ref{HL action}).\footnote{One may say that HLD action (\ref{HL action}) can couple only to the covariantly conserved matters $\hat{\nabla}_{\mu} T^{\mu \nu}_{\rm matter}=0$, when they are minimally coupled to gravity, since one can find $\hat{\nabla}_{\mu} T^{\mu \nu}_{\rm DE}=0$, independently of the matter sector~\cite{Deve:2021}.}

As a background cosmology metric, we consider the homogeneous and isotropic
FLRW metric~\cite{Frie:1922,Lema:1927}
\begin{\eq}
ds^2=-c^2dt^2+a^2(t)\left[\frac{dr^2}{1-{\cal K}r^2/R_0^2}+r^2\left(d\theta^2+\sin^2\theta
d\phi^2\right)\right]
\label{FLRW}
\end{\eq}
with the (spatial) curvature parameter ${\cal K}=+1,0,-1$ for a closed, flat, open universe, respectively, and the current curvature radius $R_0$ at $a(t_0) \equiv 1$. Assuming the perfect fluid form of matter (background) energy-momentum tensor ${\bar{T}^{\mu}}_{\nu}={\rm diag} (-\rho, p, p, p)$ with energy density $\rho$ and
pressure $p$, we obtain the Friedmann equations as
\begin{eqnarray}
\left(\frac{\dot{a}}{a}\right)^2&=&\frac{8 \pi G }{3 c^2}
(\rho_{\rm matter} +\rho_{\rm DE} )-\frac{c^2  {\cal K} }{R_0^2 a^2},  \label{FF1}\\
\frac{\ddot{a}}{a}&=&-\frac{4 \pi G }{3 c^2} [(\rho_{\rm matter} +
\rho_{\rm DE})+ 3 (p_{\rm matter} + p_{\rm DE})],\label{FF2}
\end{eqnarray}
where we introduce the energy density and pressure for the dark energy (HDE)\footnote{We follow the physical convention of Ryden~\cite{Ryde:1970,Park:2009} which disagrees with~\cite{Hora:2009}: $G_{\rm Here}=G_{\rm Horava}/c^2, \La_{\rm Here}=\La_{\rm Horava} c^2$.} as
\begin{eqnarray}
\rho_{\rm DE}&=& {\frac{3 \kappa^2 \mu^2}{8 (3 \la-1)}}
\left(
\frac{{\cal K}^2}{R_0^4 a^4}+ \frac{2 \omega {\cal K} }{R_0^2 a^2}+ \Lambda_W^2 \right), \label{rho}\\
p_{\rm DE}&=& {\frac{\kappa^2 \mu^2}{8 (3 \la-1)}}
\left( \frac{
{\cal K}^2}{R_0^4 a^4}- \frac{2 \omega {\cal K} }{R_0^2 a^2}-3 \Lambda_W^2 \right),
\label{pressure}
\end{eqnarray}
respectively, by defining the fundamental constants in GR, {\it i.e.}, speed of light $c$, Newton's constant $G$, and cosmological constant $\Lambda$ as (we have set $\la=1$)\footnote{If we consider an arbitrary $\la \neq 1$,
$\rho_{\rm DE}$ also has a  $({\dot{a}}/{a})^2$ term proportional to $(\la-1)$, but its effect is just to shift the overall factors in (\ref{rho}) and (\ref{pressure}) which corresponds to the shifts in the fundamental constraints $c, G, \La$~\cite{Park:2009} {(see Sec. V, No. 4 for further discussions)}. 
Moreover, there are some additional ambiguities in defining $\rho_{\rm DE}$ and $p_{\rm DE}$, depending on the definitions of $c$~\cite{Keha:2009,Argu:2015}. However, we have found not much difference on the main results for the background analysis~\cite{Albin:unpub}.
}
\begin{eqnarray}
c^2 = \frac{ \kappa^4 \mu^2 \Lambda_W}{ 32}, ~G=\frac{\kappa^2
c^2}{32 \pi },~ {\Lambda}=\frac{3}{2} \Lambda_{W} c^2.
\label{constant}
\end{eqnarray}
 { Note that, for the spatially-flat universe with ${\cal K} = 0$, all the contributions from
the higher-derivative terms disappear and we recover the same background cosmology as in
GR, which means a return to the flat LCDM model.
On the other hand}, it is important to note that the energy-momentum tensor of dark energy has the perfect fluid form\footnote{This is essentially due to the property of background FLRW metric. If we consider a spatially anisotropic (Bianchi) universe, we will have anisotropic pressures.}
\begin{\eq}
\bar{T}^{\mu}_{\nu ~{\rm DE}}={\rm diag} (-\rho_{\rm DE}, p_{\rm DE}, p_{\rm DE}, p_{\rm DE})
\label{EM_DE}
\end{\eq}
and satisfies the covariant conservation law
\begin{\eq}
\bar{\nabla}_{\mu} \bar{T}^{\mu \nu}_{\rm DE}=\dot{\rho}_{\rm DE}+3 H \left(\rho_{\rm DE}+p_{\rm DE}\right)=0
\end{\eq}
with the Hubble parameter $H(t) \equiv \dot{a}/a$, consistently with (\ref{Einstein_eq}), where $\bar{\nabla}_{\mu}$ denotes the covariant derivative with respect to the background metric (\ref{FLRW}).

\begin{figure}
\includegraphics[width=6cm,keepaspectratio]{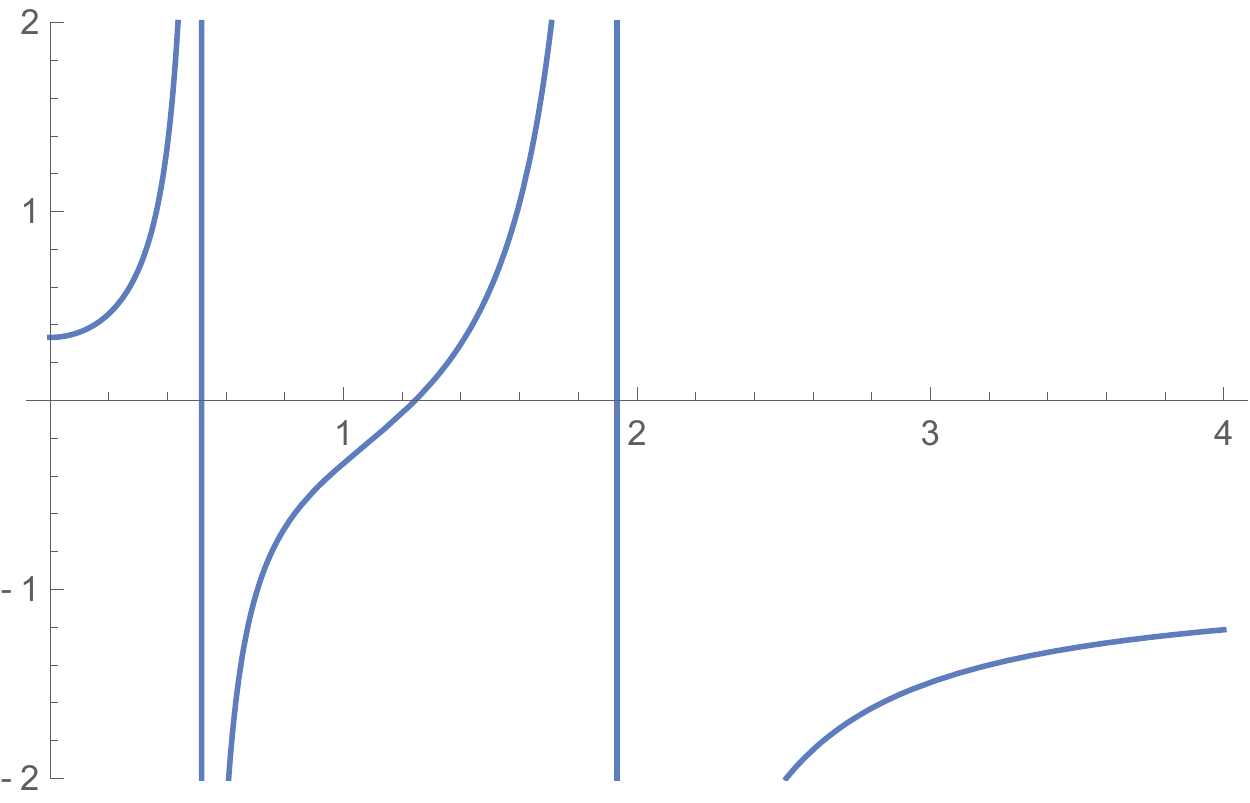}~~
\includegraphics[width=6cm,keepaspectratio]{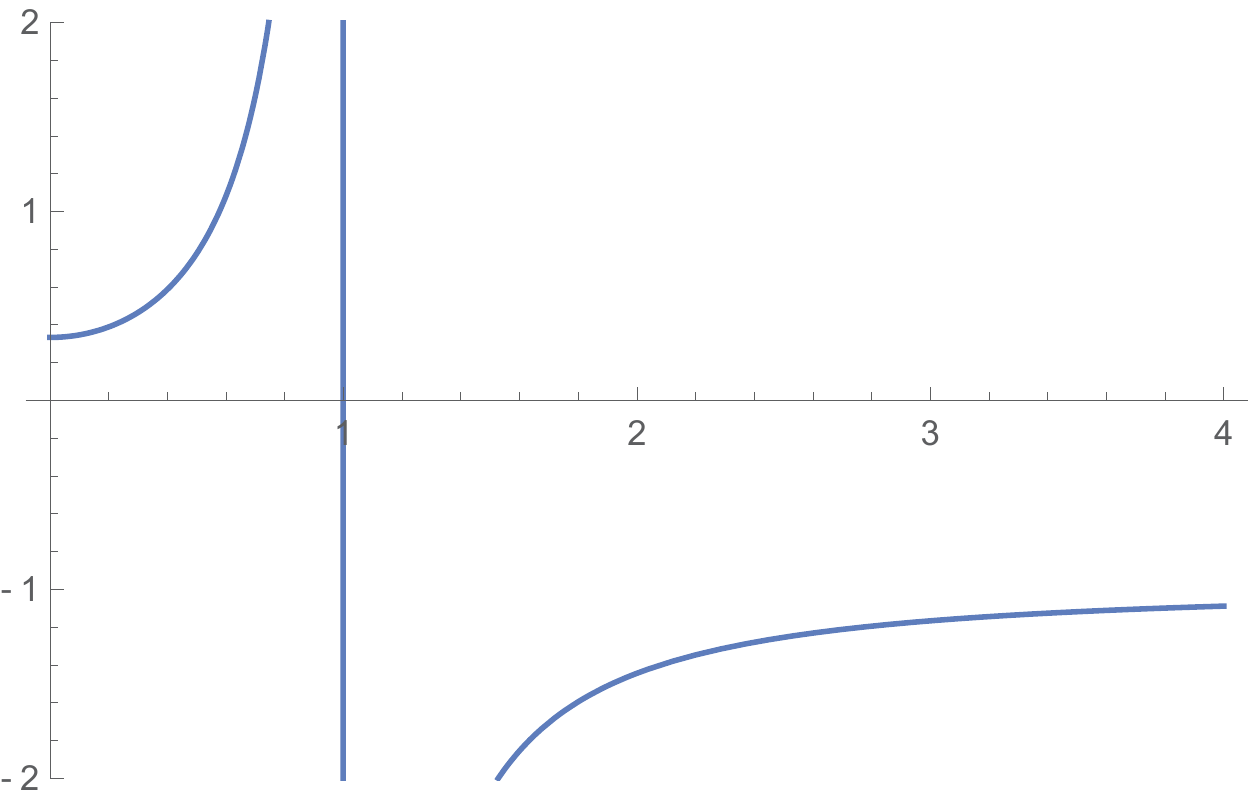}
\includegraphics[width=6cm,keepaspectratio]{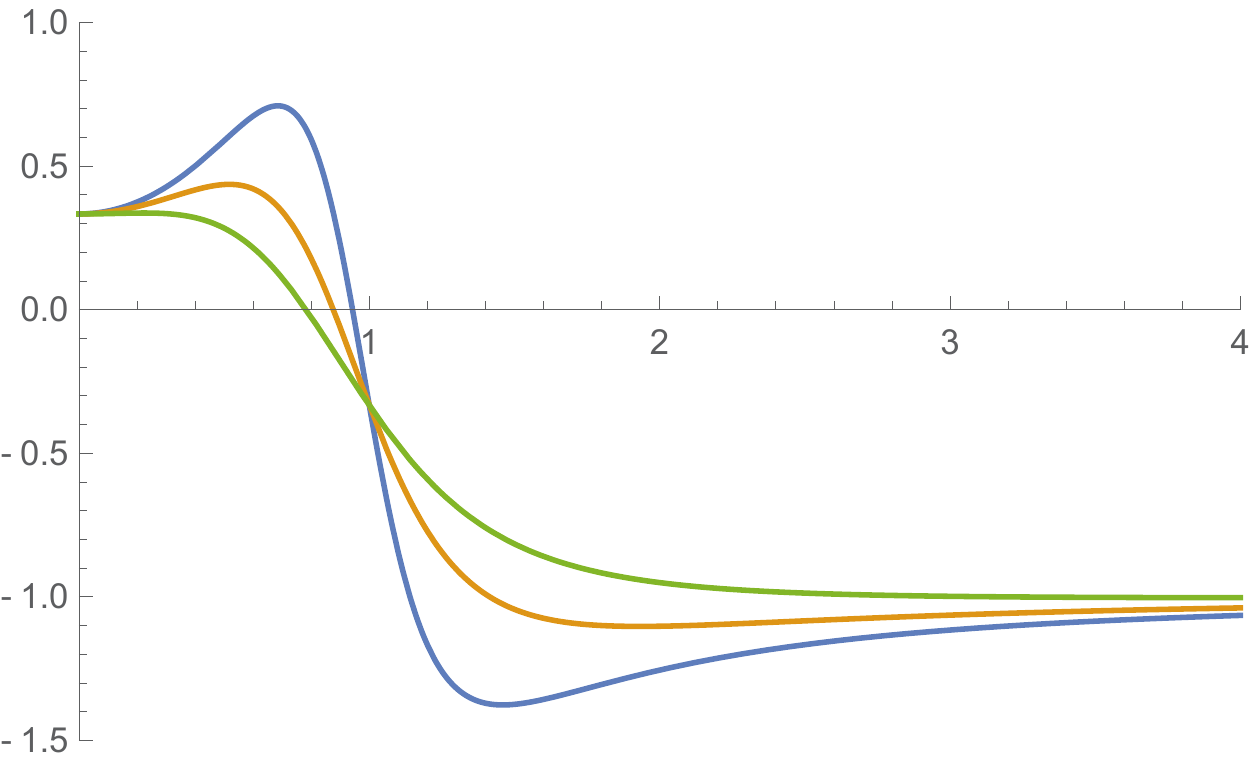}~~
\includegraphics[width=6cm,keepaspectratio]{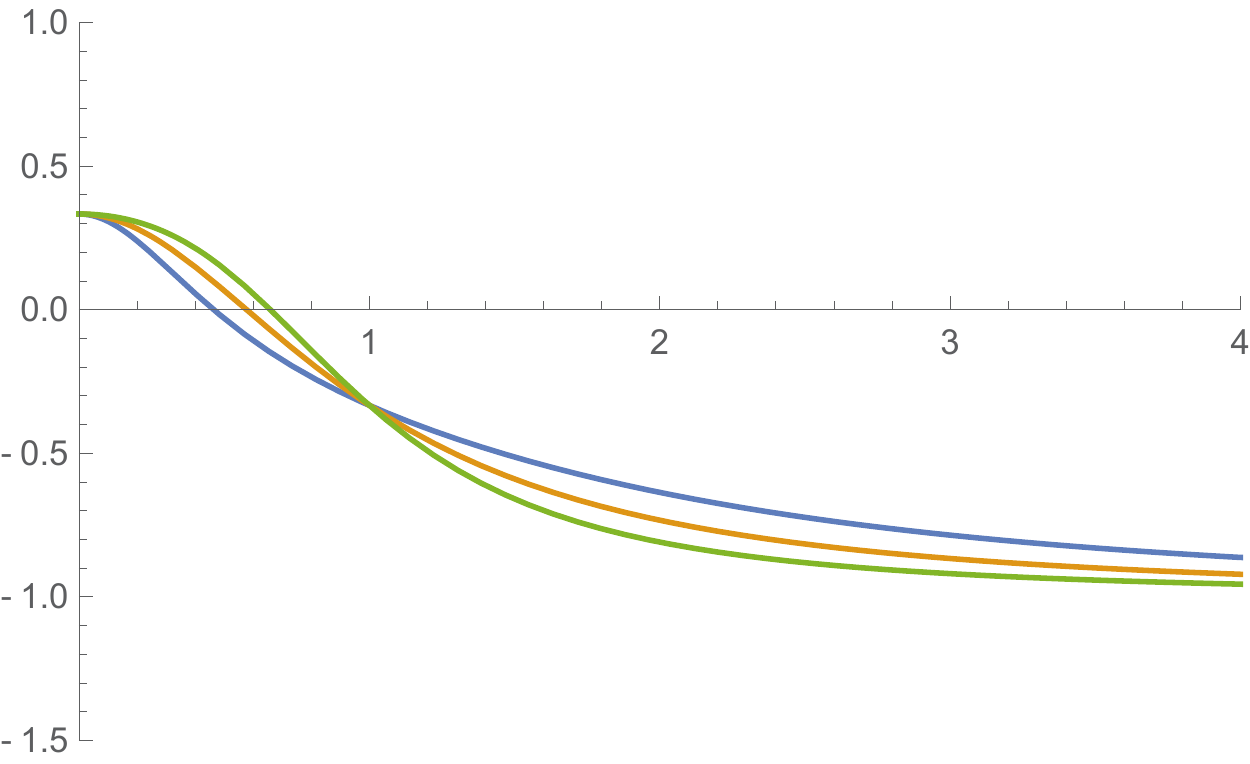}
\caption{Plots of $w_{\rm DE}$ vs. $a(t)$ for (a) $\bar{\omega}^2 > \bar{\Lambda}_W^2,~ \bar{\om}{\cal K} <0$ (top left), (b) $\bar{\omega}^2 = \bar{\Lambda}_W^2,~ \bar{\om}{\cal K} <0$ (top right), (c) $\bar{\omega}^2 < \bar{\Lambda}_W^2,~  \bar{\om}{\cal K}<0$ (bottom left), (d) and $\bar{\om}{\cal K} >0$ (bottom right). Here, we consider $\bar{\om}=-2$ (top left), $\bar{\om}=-1$ (top right), $\bar{\om}=-1/1.3, -1/2, -1/10$ (bottom left),  $\bar{\om}=2, 1, 1/2$ (bottom right) with ${\cal K}=+1$ (closed universe). Note that there is a degeneracy between ($\bar{\om}, {\cal K}$) $\lra$ $(-\bar{\om}, -{\cal K})$ in $w_{\rm DE}$ of (\ref{EOS}).} \label{fig:EOS}
\end{figure}

The equation of state (EoS) parameter is given by
\begin{eqnarray} \label{wDE}
w_{\rm DE}=\frac{p_{\rm DE}}{\rho_{\rm DE}}=\left( \frac{{\cal K}^2- 2
\bar{\omega} {\cal K}  a^2-3 \bar{\Lambda}_W^2 a^4}{3{\cal K}^2+ 6  \bar{\omega} {\cal K}  a^2+3
\bar{\Lambda}_W^2  a^4} \right) \label{EOS},
\end{eqnarray}
where we introduce $\bar{\omega} \equiv \omega
R_0^2,~\bar{\Lambda}_W=\Lambda_W R_0^2$ for convenience. For a flat universe, ${\cal K}=0$, we have $w_{\rm DE}=-1$, which just corresponds to the cosmological constant. However, for any non-flat universe, EoS interpolates from $w_{\rm DE}=1/3$ ({\it i.e.}, radiation-like) in the UV limit ($a = 0$) to $w_{\rm DE}=-1$ in the IR limit ($a = \infty$), but other detailed evolution depends on the parameters $\bar{\om}, {\cal K},  \bar{\La}_W$ (Fig. 1). In other words, ``any deviation from the cosmological constant case, $w_{\rm DE}=-1$, in the observational data (see~\cite{DES:2022} for a recent result), is an indication of a non-flat universe in our context''. In this sense, \Ho~gravity can be a ``natural laboratory'' to test a non-flat cosmology. However, we remark that the observation of $w_{\rm DE}$ only or its derived quantities can not uniquely determine the theory parameters $\bar{\omega}, {\cal K}$ due to a degeneracy between ($\bar{\om}, {\cal K}$) $\lra$ $(-\bar{\om}, -{\cal K})$ in (\ref{EOS})~\cite{Park:2009}.

\subsection{Cosmological Perturbations on a Non-Flat Background}

We consider the perturbed metric as
\begin{\eq}
N=a(\eta) [1+ {\cal A} (\eta, {\bf x})],~N_i=a^2(\eta) {\cal B}_i (\eta,{\bf x}), ~
g_{ij}=a^2(\eta) [ \ga_{ij}+h_{ij}(\eta,{\bf x})],
\end{\eq}
where $\eta$ is the conformal time, defined by $d \eta=dt/a$, and $\ga_{ij}$ is the metric for a non-flat $3$-sphere, $\ga_{ij}=(1+\bar{\cal K} r^2/4)^{-2} \de_{ij}$, by which the Latin indices $(i,j,\cdots)$ are raised with $\bar{\cal K} \equiv {\cal K}/R^2_0$.

Following the usual scalar-vector-tensor decomposition, we also consider
\begin{\eq}
\label{decomposition_0i}
{\cal B}_i &=&  D_i {\cal B} + S_i \, ,
\no \\
\label{decomposition_ij}
h_{ij} &=& 2 {\cal R} \ga_{ij}+ D_i D_j {\cal E} + D_{(i} F_{j)} + {H}_{ij} \, ,
\end{\eq}
where $D_i$ is the covariant derivative with respect to $\ga_{ij}$, and
$S_i$, $F_i$, ${H}_{ij}$ are transverse vectors and transverse-traceless
tensor, respectively, {\it i.e.,}
\begin{equation}
D_i S^i = D_i F^i = H^i_i = D_i {H}^i_j = 0 \, .
\end{equation}
Here, we note that the background FLRW metric is the {\it projectable} form, {\it i.e.}, $N=N(\eta)$~\cite{Hora:2009}, but the perturbed metric is {\it non-projectable} with an arbitrary space and time dependence, $N=N(\eta, {\bf x})$, generally {so that one can obtain the {\it local} Hamiltonian constraint by varying $N=N(\eta, {\bf x})$~\cite{Park:2009gf,Shin:2017}. This is a crucial difference from the projectable model~\cite{Hora:2009,Muko:2009}, 
where just the {\it global} Hamiltonian constraint exists so that there is an immovable gap between projetable model and GR, having the local Hamiltonian constraint, and that is the origin of their possessing extra graviton mode \footnote{{Of course, the local Hamiltonian constraint does not mean the symmetry generator as in GR. Actually, it is the {\it second-class} constraint at the fully non-linear level for either the standard non-projectable HLD action (\ref{HL action})~\cite{Deve:2021} or the $a_i$-extended non-projectable model~\cite{Bellorin:2011ff} and there are more (the second-class and the first-class) constraints than GR. However, it is important to note that, in contrast to the $a_i$-extended model, there is a case (called Case {\bf A}) where the physical degree of freedom is the same as in GR, even at the fully non-linear level for the non-projectable HLD model. It is still an open problem whether our cosmology corresponds to that case or not.}}.} 

Now, let us consider a perfect fluid of density $\rho$ and pressure $p$ with the energy-momentum tensor
\begin{\eq}
T_{\mu \nu}=(\rho+p) u_{\mu} u_{\nu} + p g_{\mu \nu}
\end{\eq}
with the $4$-velocity of a comoving observer, $u_{\m} u^{\m}=-1$, as in the matter and dark energy (\ref{EM_DE}) for the background cosmology. The general form of the perturbed energy-momentum tensor at the linear order can be written~\cite{Koda:1984,Pete:2013},
\begin{\eq}
\de T_{\mu \nu}=(\de \rho+ \de p) u_{\mu} u_{\nu} + \de p g_{\mu \nu}+2 (\rho+ p) u_{(\mu} \de u_{\nu)}+p \de g_{\mu \nu}+a^2 \Si_{\mu \nu},
\label{de T}
\end{\eq}
where non-perturbed quantities $u_{\mu}, g_{\mu \nu}, p, \rho$ denote the background objects, $\de u^{\mu}=a^{-1}(-{\cal A}, v^i)$, and $\Si_{\mu \nu}$ is the anisotropic stress tensor which represents a non-perfect fluid perturbation, $\Si_{ij}=T_{ij}-(1/3){T^k}_k  \ga_{ij} , {\Si^{i}}_i=\Si_{0\mu}=0$, absorbing its trace by the pressure $p$.

In component form, the perturbations read
\begin{\eq}
\de T_{00}&=&a^2 \rho  \left(\de_{\rho}+ 2 {\cal A}\right), \no \\
\de T_{0i}&=&-a^2(\rho+p)  (v_i+ {\cal B}_i), \no \\
\de T_{ij}&=&a^2 \left(p h_{ij} +\ga_{ij} \de {p}+ \Si_{ij}\right),
\label{de T_comp}
\end{\eq}
where $\de_{\rho}\equiv \de \rho/\rho$ is the density contrast.

In the cosmological perturbation analysis, a proper choice of gauge is
useful, depending on the situation being studied. In \Ho~ gravity,
the ``apparent'' action symmetry is the ``foliation-preserving'' {\it Diff} symmetry (${\it Diff_{\cal F}}$) under the coordinate transformation $\de \eta=-f(\eta), \de x^i =-\xi^i (\eta, {\bf x})$, and hence not all the gauges in GR may be allowed. In this paper, we consider the synchronous gauge, which is one of the allowed gauges in \Ho~ gravity, where we
set ${\cal A}={\cal B}_i=0$, since ${\cal A}$ and ${\cal B}_i$ are the
Lagrange multipliers and can be integrated out in the Hamiltonian-reduction (or Faddeev-Jackiw)
method~\cite{Shin:2017}. Actually, in \Ho~ gravity, this is a natural gauge without much loss of generality of ${\it Diff_{\cal F}}$, since the residual symmetry of the synchronous gauge corresponds to a
${\it Diff_{\cal F}}$ with $\de \eta=-f(\eta)$, $\de x^i =-\xi^i ( {\bf x})$.

In the synchronous gauge, the equations for the perturbed fluid are given by the perturbed conservation equations, $\de \hat{\nabla}_{\mu} {T^{\mu}}_{\nu}=0$ (in momentum space)~\cite{Koda:1984,Ma:1995ey,Hu:1998kj},
\begin{\eq}
{\de_{\rho}}'&=&-3 {\cal H} \left(\f{\de p}{\de \rho}-w \right) \de_{\rho} -(1+w) \left(\theta+\f{h'}{2}\right),  \label{pert_Eq1}\\
{\theta}'&=&-{\cal H} (1-3 w) \theta-\f{w'}{1+w} \theta +\f{\de p/\de \rho}{1+w} k^2 \de_{\rho}-k^2 \sigma,
\label{pert_Eq2}
\end{\eq}
where the prime $(')$ denotes the derivative with respect to conformal time $\eta$, ${\cal H}\equiv (a'/a)$, $\theta \equiv i k^i v_i, h\equiv {h^i}_i, \si \equiv -(\hat{k}_i \hat{k}_j-\de_{ij}/3) {\Si^i}_j/(\rho+p)$, and $k^i$ is the comoving wavevector in a {\it non-flat space} generally, defined by
\begin{\eq}
{D}^i A_{jk \cdots}=i k^i A_{jk \cdots},~{D}^i {D}_i A_{jk \cdots}=- k^2 A_{jk \cdots}
\end{\eq}
with $k^2 \geq |{\bar {\cal K}}|$ for ${\bar {\cal K}} \leq 0$, or $k^2=l(l+4) {\bar {\cal K}}~ (l=0,1,2, \cdots)$ for ${\bar {\cal K}}>0$ with the appropriate eigenfunctions (harmonics) $A_{jk \cdots}$~\cite{Vile:1964,Hu:1997mn}.

In terms of the gauge invariant, {\it i.e.} physical, sound speed $\hat{c}_s$ in the rest frame~\cite{Bard:1980,Koda:1984,Hu:1998kj},
\begin{\eq}
\f{\de p}{\de \rho}=\hat{c}^2_s+\left[3 {\cal H} (\hat{c}^2_s-w)+w' \right] \f{\rho}{\de \rho} \f{\theta}{k^2},
\end{\eq}
the perturbation equations (\ref{pert_Eq1}), (\ref{pert_Eq2}) reduce to
\begin{\eq}
{\de_{\rho}}'&=&-3 {\cal H} (\hat{c}^2_s-w ) \de_{\rho}+3 {\cal H} \left[3 {\cal H} (1+w)(\hat{c}^2_s-w)+w' \right]\f{\theta}{k^2} -(1+w) \left(\theta+\f{h'}{2}\right),  \label{pert_Eq1b}\\
{\theta}'&=&-{\cal H} (1-3 \hat{c}^2_s) \theta+\f{\hat{c}^2_s}{1+w} k^2 \de_{\rho}-k^2 \sigma,
\label{pert_Eq2b}
\end{\eq}
where we have used the definition of the ``adiabatic'' sound speed,
\begin{\eq}
c^2_a \equiv
\f{p'}{\rho'}=w-\f{w'}{3 {\cal H} (1+w)}
\label{c_a}
\end{\eq}
for the {\it adiabatic} perturbations, $\de_{a} p\equiv c^2_a \de_{a} \rho$.
Here, note that the adiabatic sound speed $c_a$ is determined by the background quantities only
and can be infinite at $w=-1$, {\it i.e.}, cosmological constant,
unless $w'=0$, or even negative depending on $w$: it just represents $\rho'=0$ for the former, or even $\rho'<0$
for the latter while $p'>0$
(or vice versa).

So far, we have not specified any {particular} fluid and the equations are valid for any ``uncoupled'' fluid. Now, by introducing dust matter (non-relativistic baryonic matter and (non-baryonic) cold dark matter with $p_m=0$) and radiation
(ultra-relativistic matter with $p_r=\rho_r/3$), which satisfy the continuity equations
$\hat{\nabla}_{\mu} T^{\mu \nu}_{(i)}=0~(i=m, r)$, we can write the Friedmann equation (\ref{FF1}) as
\begin{equation}\label{H_O_eq}
    \left(\frac{H}{H_0}\right)^2 = \Omega_r a^{-4}+\Omega_m a^{-3}  + \Omega_k a^{-2} + \Omega_{\rm DE}(a).
\end{equation}
Here, we define the canonical density parameters at the current epoch ($a_0=1$) as\footnote{We adopt the convention $\Om_{i}$ for the current values and $\Om_{i} (a)$ for the fully time-dependent values.}
\begin{\eq}
\Omega_m \equiv \be \f{ \rho^0_m}{3H_0^2},~ \Omega_r\equiv \be \f{\rho^0_r}{3H_0^2},~ \Omega_k\equiv-\ga \f{{\cal K}}{H_0^2R_0^2}, ~ \Omega_\Lambda\equiv\ga \f{\Lambda_W}{2H_0^2},~ \Omega_\omega\equiv\ga \f{\omega}{2H_0^2}
\end{\eq}
\\
with positive parameters $\be \equiv  \kappa^2/2(3\lambda-1)$, $\ga \equiv \kappa^4 \mu^2 \La_W/8(3\lambda-1)^2$ and introduce the (dynamical) dark-energy (HDE) component~\cite{Nils:2021} as
\begin{equation}
    \Omega_{\rm DE}(a) \equiv \left(\frac{\Omega_k^2}{4\Omega_\Lambda}\right) a^{-4}
    -\left(\frac{\Omega_k \Omega_\omega }{\Omega_\Lambda}\right)a^{-2} + \Omega_\Lambda,
    \label{DE}
\end{equation}
which includes the {\it dark radiation} ($\sim a^{-4}$) and {\it dark curvature} ($\sim a^{-2}$) components as well as the cosmological constant component $\Omega_\Lambda$. The Friedmann equation (\ref{H_O_eq}) is now given by
\begin{equation}\label{Model_B}
    \left(\frac{H}{H_0}\right)^2 = \left( \Omega_r + \frac{\Omega_k^2}{4\Omega_\Lambda} \right) a^{-4}+\Omega_m a^{-3}  + \left(1-\frac{\Omega_\omega }{\Omega_\Lambda}\right) \Omega_k a^{-2} + \Omega_{\La}.
\end{equation}
The EoS parameter (\ref{EOS}) can be written as
\begin{\eq}
w_{\rm DE}(a)=\f{-12~ \Om^2_{\La} a^4 +4 \Om_k \Om_\om a^2 +\Om_k^2}{12~ \Om^2_{\La} a^4 -12 \Om_k \Om_\om a^2 +3\Om_k^2}.
\label{EOS_O}
\end{\eq}
Then, from (\ref{c_a}), the adiabatic sound speed can be obtained as
\begin{\eq}
c^2_a=\f{{\cal K}+\bar{\om} a^2 }{3({\cal K}-\bar{\om} a^2 )}=\f{-2 \Om_{\om} a^2 +\Om_k}{6 \Om_{\om} a^2 +3\Om_k},
\end{\eq}
which goes to
$c^2_a=1/3$ in the UV limit (when ${\cal K} \neq 0$) and $c^2_a=-1/3$ in the IR limit, but the evolution details depend on $\bar{\om}, {\cal K}$ or $\Om_{\om}, \Om_k$ (Fig. 2). It is interesting to note that the cosmological constant parts of $w_{\rm DE}$ in (\ref{EOS_O}) do not contribute to $c_a$.

\begin{figure}
\includegraphics[width=8cm,keepaspectratio]{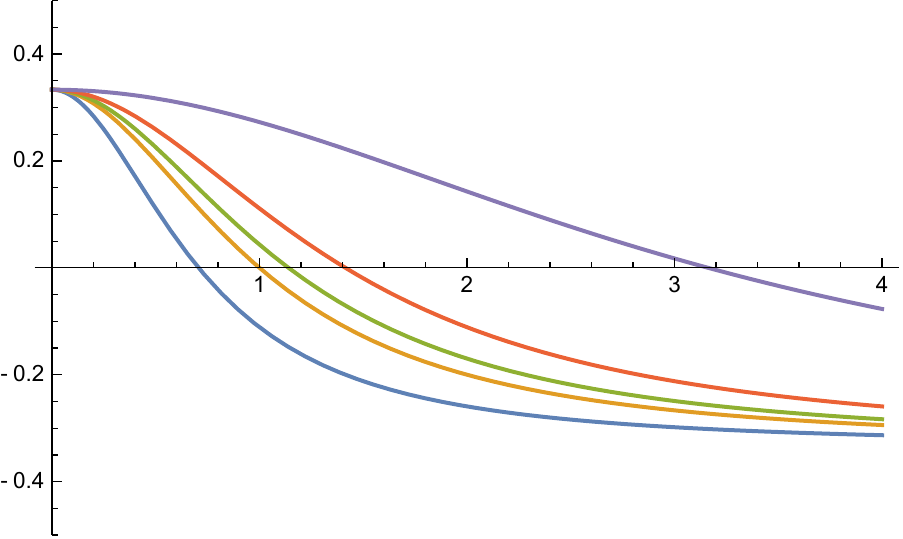}~~
\includegraphics[width=8cm,keepaspectratio]{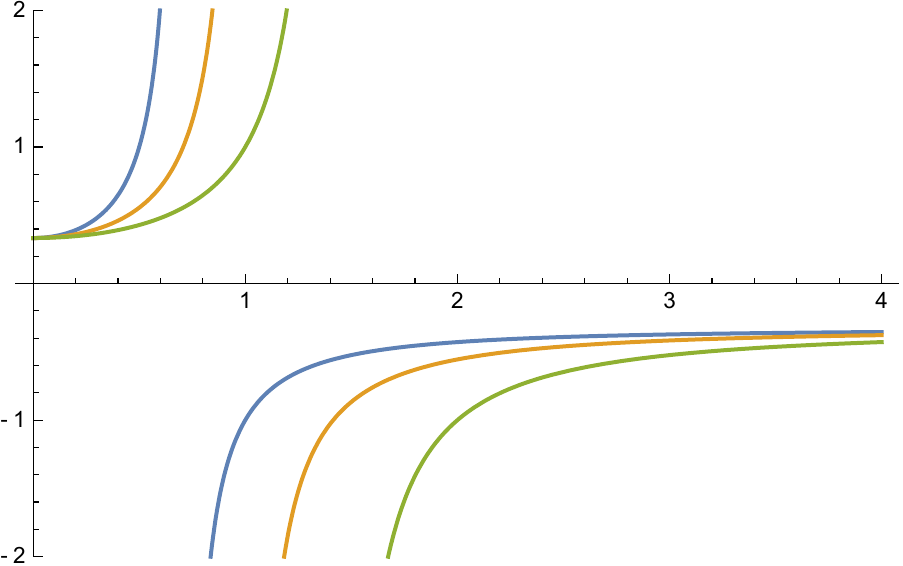}
\caption{Plots of $c^2_a$ vs. $a(t)$ for $\bar{\om}{\cal K} <0$ (left) and $\bar{\om}{\cal K} >0$ (right). Here, we consider the same values of parameters as in Fig. 1: $\bar{\om}=-2,-1,-1/1.3, -1/2, -1/10$ (left, bottom to top) correspond to the first three cases in Fig. 1, whereas $\bar{\om}=2, 1, 1/2$ (right, top to bottom)
correspond to the last case in Fig. 1, with ${\cal K}=+1$ (or $\bar{\om}$ $\ra$ $-\bar{\om}$ with ${\cal K}=-1$). Note that $c^2_a$ evolves from $c^2_a=1/3$ in the UV limit to $c^2_a=-1/3$ in the IR limit, but other intermediate details depend on $\bar{\om}$ and ${\cal K}$.} \label{fig:cs2}
\end{figure}

\subsection{Comparison to  Chevallier-Polarski-Linder (CPL) Type Parametrization}
\begin{figure}
\includegraphics[width=8cm,keepaspectratio]{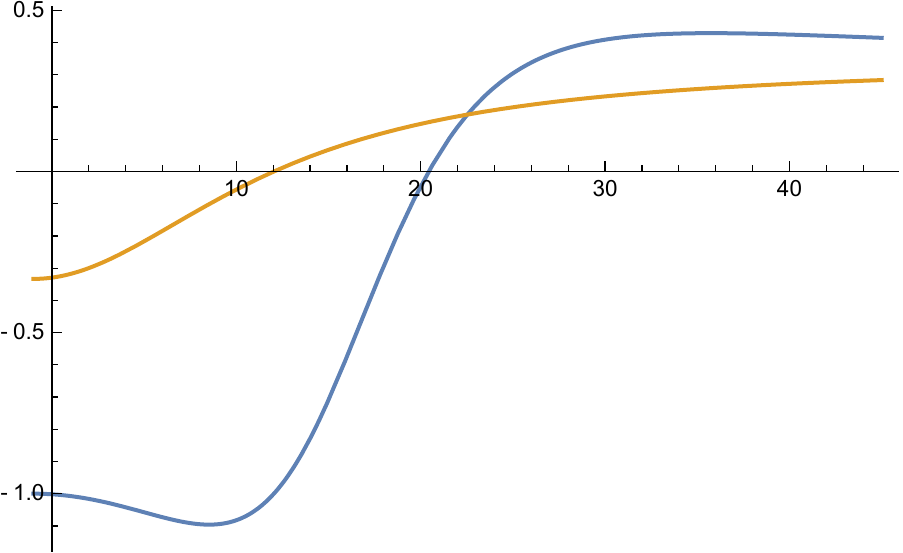}
\caption{Plots of $w_{\rm DE}$ (blue), $c^2_a$ (yellow) vs. $z=-1+1/a$ for our preferred case $\bar{\om}{\cal K} <0$ or $\Om_{\om} \Om_k>0$.
Here, we take $\Om_m=0.306, \Om_r=8.64 \times 10^{-5}, \Om_k=-0.004, \Om_\La= 0.695, H_0=69.53, \Om_\om=-0.34$ from the previous background analysis, but with our preferred signature of $\Om_\om<0$~\cite{Nils:2021}.} \label{fig:EOS_preferred}
\end{figure}

\begin{figure}
\includegraphics[width=7cm,keepaspectratio]{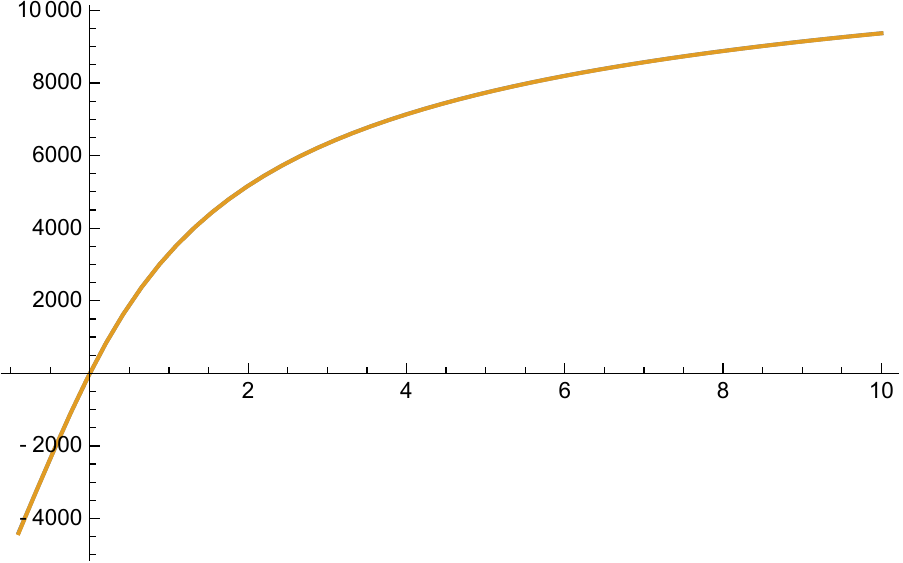}\\
\includegraphics[width=7cm,keepaspectratio]{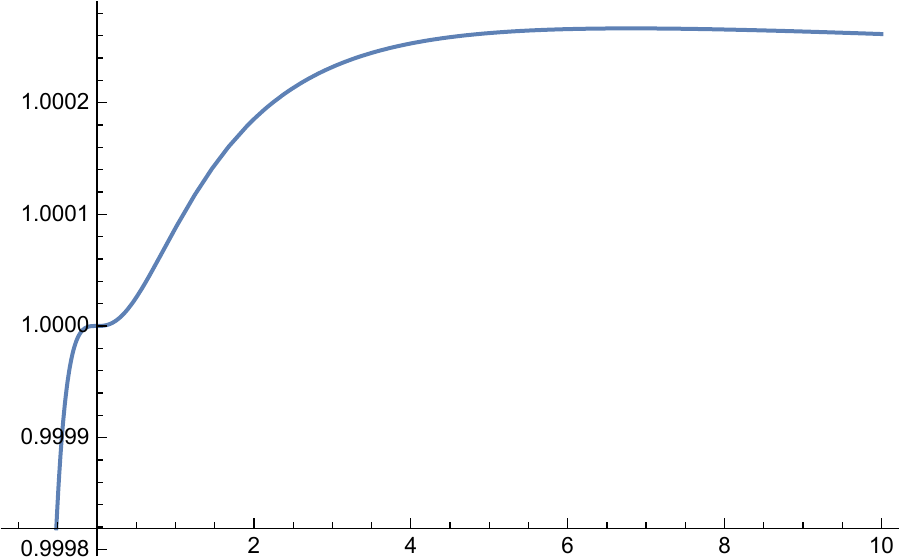}~~
\includegraphics[width=7cm,keepaspectratio]{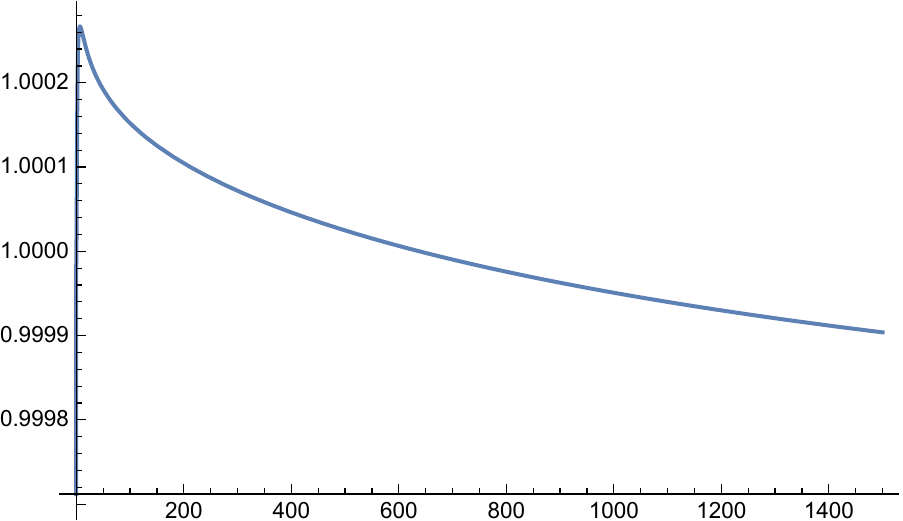}
\caption{Comparison of the comoving angular-diameter distance $d^c_A(z)$ between our dark energy model for the preferred case in Fig. 3 and the corresponding CPL model with $w_0=-1.002, w_a=-0.004$. The top plot shows the distances for the two cases with quite a good fit (the two curves are almost coincident). The bottom plots show the ratio of our preferred case vs. the CPL model which provides the robustness of our dynamical dark energy model, even up to the CMB distance.}  \label{fig:distance}
\end{figure}

Our dark energy model described by $w_{\rm DE}$ (\ref{wDE}) may have very rapid evolution or fluctuations with a phantom crossing at $w_{\rm DE}=-1$ as can be seen in Fig.~\ref{fig:EOS_preferred}. This might raise some questions on the relation of our dynamical dark energy model to the Chevallier-Polarski-Linder (CPL) type parametrization~\cite{Chev:2000,Lind:2002},
\begin{\eq}
w(a)= w_0+w_a (1-a)+w_b (1-a)^2+\cdots,
\label{CPL}
\end{\eq}
which is smoothly evolving but provides an excellent fit (at about $0.1 \%$ level in observables).

To that end, we consider the comoving angular-diameter distance
\begin{\eq} \label{comovangdiam}
   d^c_A (z) =\f{c}{H_0} \f{1}{\sqrt{\Om_k}}~ {\rm Sinh} \left[ \sqrt{\Om_k}  \int_0^z \frac{dz'}{H (z')/H_0} \right]
\end{\eq}
for the redshift $z=-1+1/a$.
Here, the Hubble parameter $H(z)$ can be written as
\begin{\eq}
\label{H_O_eq_omega}
    \left(\frac{H(z)}{H_0}\right)^2 = \Omega_r (z+1)^{4}+\Omega_m (z+1)^{3}  + \Omega_k (z+1)^{2} + \Omega_{\rm DE}\cdot{\rm exp}\left(3 \int^z_0 \f{1+w_{\rm DE}(z')}{1+z'}dz'\right)
\end{\eq}
from the Friedmann equation (\ref{H_O_eq}) and the dark energy density parameter which solves the background conservation law (14),
\begin{\eq}
\Omega_{\rm DE}(z)&=&\Omega_{\rm DE}\cdot {\rm exp}\left[3 \int^z_0 \f{1+w_{\rm DE}(z')}{1+z'}dz'\right]  \\
&=&\Omega_{\rm DE}\cdot {\rm exp}\left[3 (1+w_0+w_a) z(z+2)- \f{3w_a z}{z+1}
+3 w_b \left( ln(z+1)-\f{z(z+2)}{2 (z+1)^2} \right)+\cdots \right] \no
\label{Omega_sol}
\end{\eq}
in the CPL-type parametrization (\ref{CPL}). Now, in order to see the goodness of the {\it standard} CPL model with the first two parameters $w_0, w_a$, we compare the comoving angular-diameter distance (\ref{comovangdiam}) between our model with the full $w_{\rm DE}$ (\ref{wDE}) and the CPL model with 
\begin{\eq}
w_{0}=\f{-12~ \Om^2_{\La} +4 \Om_k \Om_\om +\Om_k^2}{3 ~(4 \Om^2_{\La}  -4 \Om_k \Om_\om  +\Om_k^2)},~
w_a= \frac{16~\Omega_k(4\Omega_\omega\Omega_\Lambda^2
-4\Omega_k\Omega_\Lambda^2+\Omega_k^2\Omega_\omega)}{3~(4 \Om^2_{\La}  -4 \Om_k \Om_\La  +\Om_k^2)^2},
\label{EOS_CPL}
\end{\eq}
which can be obtained by expanding (\ref{wDE}) near the current epoch $a=1$.
Our result in Fig.~\ref{fig:distance} shows that, even for the rapidly evolving case of dark energy in Fig.~\ref{fig:EOS_preferred}, the agreement is better than $0.03 \%$ at all redshifts ($0.006 \%$ at the CMB distance $z\approx 1100$), which is sufficient for the current precision of data. From the fact that the comoving angular-diameter distance enters the CMB distance to the surface of last scattering, BAO, and Supernovae observations, our result supports the robustness of our dynamical dark energy model (HDE) in comparison to the standard CPL model.\footnote{This circumstance is quite similar to the vacuum metamorphosis (VM) model~\cite{DiVa:2017,DiValentino:2020kha}. In fact, even the asymptotic values of its equation of state at UV and IR are the same and its rapidly evolving behaviour is quite similar. Understanding the physical relevance of the VM model to our dark energy model would be interesting.}

\subsection{Assumptions and Initial Conditions}

We have introduced the dark energy fluid which is not interacting with other matter and radiation, by its purely gravitational origin. Their perturbation equations are very general and includes anisotropic stress $\s$, time-varying $w_{\rm DE}$, and arbitrary rest-frame sound speed $\hat{c}_s$.
In the fluid approach, we have identified the background quantities $T^{\mu \nu}$ but have not specified other details about the perturbed quantities $\de T^{\mu \nu}$, coming from the Lorentz-violating higher-derivative terms in \Ho~gravity. The only model dependence enters in the background $T^{\mu \nu}$ or the EoS parameter $w_{\rm DE}$ and all the other perturbation analysis can be very general, {\it i.e.} model independent. However, from Eq.~(6), it might have its own limitation of validity by neglecting the genuine UV, {\it i.e.} deep subhorizon $(k \gg {\cal H})$, effects at the primordial universe: (6) assumes implicitly the same order of the second-order derivatives of gravitational perturbations $\de G^{\mu \nu}$ as the fluid perturbations $\de T^{\mu \nu}$, which contains the higher-order derivatives of gravitational perturbations $\de T^{\mu \nu}_{\rm{DE}}$ in addition to the conventional matter perturbations $\de T^{\mu \nu}_{\rm{matter}}$ in our \Ho~gravity model. Actually, (6) corresponds to a coarse-graining of the arbitrary perturbations and selects the spatially-slowly-varying and smooth gravity perturbations, while the background  $T^{\mu \nu}$ may allow ``rapidly-varying'' $w_{\rm DE}$ due to the UV effect in a non-flat universe.

The physical sound speed of dark-energy perturbations $\hat{c}_s$ can be arbitrary, but for the data analysis in the following sections, we will consider a {\it constant} $\hat{c}_s$ as a free parameter for simplicity; however, from the perturbation equations (26), (27), $\hat{c}^2_s$ can not be arbitrarily large for the stability of perturbations. Moreover, the anisotropic stress $\s$ is an important parameter that characterizes the perturbations~\cite{Hu:1998kj} but constraining it seems to be difficult with the current precision~\cite{Yang:2020}. Therefore, we assume the absence of anisotropic stress $\s$ in the following analysis.

For the initial conditions of dark energy perturbations in the early universe, {\it i.e.}, during the radiation-dominated era where the perturbations are outside the (Hubble) horizon, we will consider the {\it adiabatic} initial perturbations of dark energy as~\cite{Koiv:2005},
\begin{\eq}
\de_{\rm DE}=\left(\f{1+w_\ga}{1+w_{\rm DE}} \right) \de_\ga,
~\theta_{\rm DE}=\theta_{\ga},
\end{\eq}
where $\de_{\ga},\theta_{\ga}$ are the initial perturbations of photons with $w_{\ga}=1/3$. Here, the velocity condition $\theta_{\rm DE}=\theta_{\ga}$ is strictly valid for the adiabatic dark-energy fluid, {\it i.e.}, $\hat{c}^2_s=c^2_a=1/3$, at UV. However, even for $\hat{c}^2_s \neq c^2_a=1/3$, we will take the same initial condition in the following analysis since the late-time evolution is not (much) affected as far as $\hat{c}^2_s$ is inside some reasonable region~\cite{Koiv:2005}.

Finally, for the dynamical dark energy with a ``phantom crossing'', {\it i.e.}, $w_{\rm DE}=-1$, the perturbation equation (27) looks divergent, {\it i.e.}, unstable, for any non-vanishing $\hat{c}_s$. But it just means the vanishing density perturbation $\de_{\rm DE}=0$ at the instant and the equation (26) shows that later evolution may generate $\de_{\rm DE}$ again for $w'_{\rm DE} \theta>0$. In other words, the perturbation equations can be well-defined even at a phantom crossing.

\begin{figure}[h]
\begin{center}
    \includegraphics[width=0.6\textwidth]{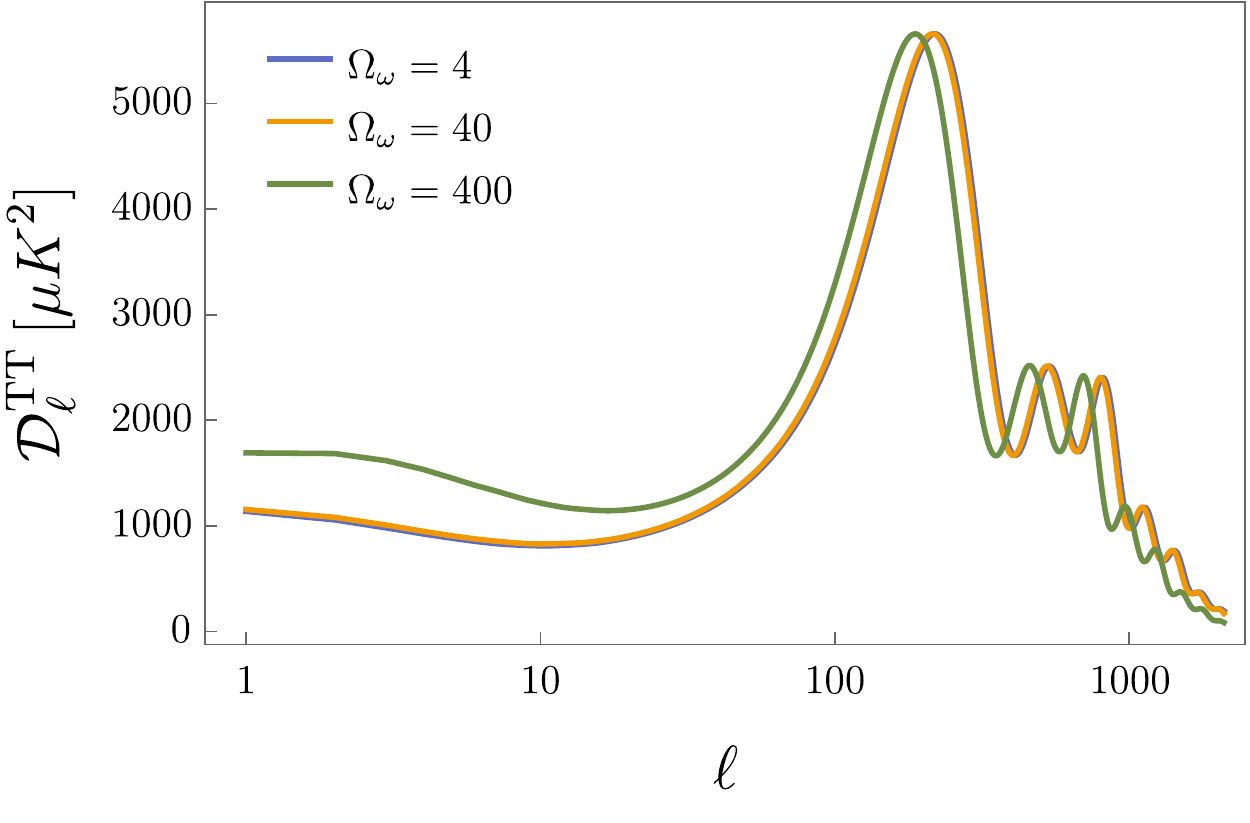}
    \caption{The CMB temperature power spectrum for different values of $\Omega_\omega$, while all the other parameters are fixed to a slightly closed universe with $\Omega_k=-0.001$. Given the symmetry for positive and negative values of $\Omega_\omega$ and $\Omega_k$ that we discuss in the text, we show only positive $\Omega_\omega$. By increasing this parameter we see a shift of the peaks towards lower multipoles, and an increase of the low-$\ell$ plateau.}
    \label{fig:TT}
\end{center}
\end{figure}

\section{Methodology and Data Sets}
\label{sec:methodology} 

In this section we list the current cosmological data sets used to constrain the HDE model:

\begin{itemize}
\item {\it Planck}: we make use of the full
Cosmic Microwave Background (CMB) temperature and polarization angular power spectra {\it plikTTTEEE+lowl+lowE} as {released} by Planck 2018 and used in~\cite{Planck:2018vyg,Planck:2019nip}.

\item {\it Lensing}: we consider the CMB lensing reconstruction likelihood from Planck 2018~\cite{Planck:2018lbu}.

\item {\it BAO}: we add the Baryon Acoustic Oscillations (BAO) distance measurements from different astronomical surveys: 6dFGS~\cite{Beutler:2011hx}, SDSS-MGS~\cite{Ross:2014qpa}, and BOSS DR12~\cite{BOSS:2016wmc}.

\item {\it Pantheon}: we include the Pantheon~\cite{Pan-STARRS1:2017jku} sample of 1048 Type Ia Supernovae.

\end{itemize}

The parameter constraints are computed by means of MCMC sampling with our modified version of the publicly available packages \texttt{CAMB}~\cite{Lewis:1999bs} and \texttt{CosmoMC}~\cite{Lewis:2002ah}.\footnote{\url{http://cosmologist.info/cosmomc/}}
The convergence diagnostic follows the Gelman and Rubin prescription~\cite{Gelman:1992zz}, which is already implemented in the Planck 2018 likelihood~\cite{Planck:2019nip}.

In Table~\ref{tab:priors} we show the {\it flat} priors adopted in the data analysis. These are: the physical density of baryons $\Omega_{b} h^2$, the physical density of cold dark matter $\Omega_{c}h^2$, the ratio of sound horizon to the angular diameter distance at recombination $\theta_{MC}$, the reionization optical depth $\tau$, the scalar spectral index $n_{s}$ and the amplitude $A_s$ of the primordial scalar power spectrum, the curvature of the universe $\Omega_k$, and the HDE
parameter $\Omega_\omega$. In particular, we analyse three cases: a closed universe where we restrict the $\Omega_k$ prior to $[-0.3,0]$, an open universe with $\Omega_k$ in $[0,0.3]$, and the full range listed in Table~\ref{tab:priors}.

In Fig.~\ref{fig:TT} we show how variations in the HDE parameter $\Omega_\omega$ affect the CMB temperature power spectrum, while all the other parameters are fixed to a slightly closed universe with $\Omega_k=-0.001$. We can see that by increasing the value of $\Omega_\omega$ we obtain a shift of the peaks towards lower multipoles, and an increase of the low-$\ell$ plateau.


\begin{table}
\begin{center}
\renewcommand{\arraystretch}{1.4}
\begin{tabular}{|c@{\hspace{1 cm}}|@{\hspace{1 cm}} c|}
\hline
Parameter                    & Prior\\
\hline\hline
$\Omega_{b} h^2$             & $[0.005,0.1]$\\
$\Omega_{c} h^2$             & $[0.001,0.99]$\\
$100\theta_{MC}$             & $[0.5,10]$\\
$\tau$                       & $[0.01,0.8]$\\
$n_s$                        & $[0.8, 1.2]$\\
$\log[10^{10}A_{s}]$         & $[1.61,3.91]$\\
$\Omega_{k}$             & $[-0.3,0.3]$\\
$\Omega_{\omega}$             & $[{-100},100]$\\
\hline
\end{tabular}
\end{center}
\caption{Flat priors adopted on the free parameters of the HDE model explored here.}
\label{tab:priors}
\end{table}


\begingroup
\squeezetable
\begin{center}
\begin{table*}[!]
\renewcommand{\arraystretch}{2}
\begin{tabular}{l|@{\hspace{0.5 cm}} ccccccccc}
\hline\hline
& Planck&Planck & Planck&Planck&Planck+Lensing \\
Parameters & &+Lensing & +BAO&+Pantheon&+BAO+Pantheon \\
\hline

$\Omega_b h^2$ & $    0.02255\pm0.00017$ & $    0.02248\pm0.00016$  & $    0.02239\pm0.00015$ & $    0.02251\pm0.00016$ & $    0.02240\pm0.00015$ \\

$\Omega_c h^2$ & $    0.1085\pm0.0048$ & $    0.1155\pm0.0034$ & $    0.1191\pm0.0021$ & $    0.1105\pm0.0044$ & $    0.1190\pm0.0021$  \\

$100\theta_{MC}$ & $    1.04130\pm0.00037$ & $    1.04112\pm0.00034$  & $    1.04100\pm0.00032$ & $    1.04124\pm0.00034$ & $    1.04100\pm0.00031$\\

$\tau$ & $    0.0517\pm0.0084$ & $    0.0517\pm0.0081$ & $    0.0551\pm0.0080$ & $    0.0525^{+0.0081}_{-0.0073}$ & $    0.0545\pm0.0076$ \\

${\rm{ln}}(10^{10} A_s)$ & $    3.032\pm0.018$ & $    3.034\pm0.017$  & $    3.044\pm0.016$ & $    3.035\pm0.017$ & $    3.043\pm0.015$ \\

$n_s$ & $    0.9729\pm0.0054$ & $    0.9696\pm0.0050$ & $    0.9668\pm0.0045$ & $    0.9716\pm0.0051$ & $    0.9670\pm0.0044$ \\

$\Omega_{k}$ & $    0.00089^{+0.00029}_{-0.00034}$ & $    0.00056^{+0.00018}_{-0.00047}$  & $    <0.000501$ & $    0.00081^{+0.00029}_{-0.00037}$ & $    <0.000464$ \\

$\Omega_{\omega}$ & $    >56.1$ & $    40^{+20}_{-30}$  & $    23\pm30$ & $    61^{+30}_{-20}$ & $    27^{+20}_{-30}$ \\

\hline

$\Omega_{\Lambda}$ & $  0.7146\pm0.0077$ & $  0.7012\pm0.0043$  & $  0.6966^{+0.0023}_{-0.0026}$ & $  0.7112\pm0.0069$ & $  0.6965^{+0.0021}_{-0.0023}$ \\

$\Omega_{m}$ & $    0.2845\pm0.0079$ & $    0.2983\pm0.0044$  & $    0.3030^{+0.0028}_{-0.0024}$ & $    0.2880\pm0.0071$ & $    0.3031^{+0.0025}_{-0.0022}$ \\

$\sigma_8$ & $    0.857^{+0.019}_{-0.014}$ & $    0.823\pm0.010$   & $    0.820^{+0.012}_{-0.014}$ & $0.852^{+0.018}_{-0.015}$ & $    0.8189\pm0.0092$ \\

$S_8$ & $    0.834\pm0.011$ & $    0.8210\pm0.0090$ & $    0.824\pm0.012$ & $    0.834\pm0.011$ & $    0.8231\pm0.0088$ \\

$H_0~[km/s/Mpc]$ & $   68.04\pm0.44$ & $   68.18\pm0.43$ & $   68.49\pm0.39$ & $   68.13\pm0.43$ & $   68.45\pm0.36$ \\

$r_{\rm{drag}}$ & $  147.29\pm0.31$ & $  147.30\pm0.32$  & $  147.13\pm0.31$ & $  147.26\pm0.31$ & $  147.15\pm0.30$ \\

\hline

$\chi^2_{\rm{bestfit}}$ & $  2755.4$ & $  2771.4$  & $  2770.5$ & $  3792.4$ & $  3815.4$ \\

$\Delta \chi^2_{\rm{bestfit}}$ & $ -10.4$ & $  -3.2$  & $ -1.4 $ & $  -8.7 $ & $ -1.6 $ \\


\hline\hline
\end{tabular}
\caption{Parameter constraints at 68\% CL for an {\it open} universe on the independent (above the line) and dependent (below the line) parameters, together with the best-fit $\chi^2$ and its difference from the flat LCDM model. 
}
\label{tab:open}
\end{table*}
\end{center}
\endgroup


\begin{figure*}
\centering
\includegraphics[width=\textwidth]{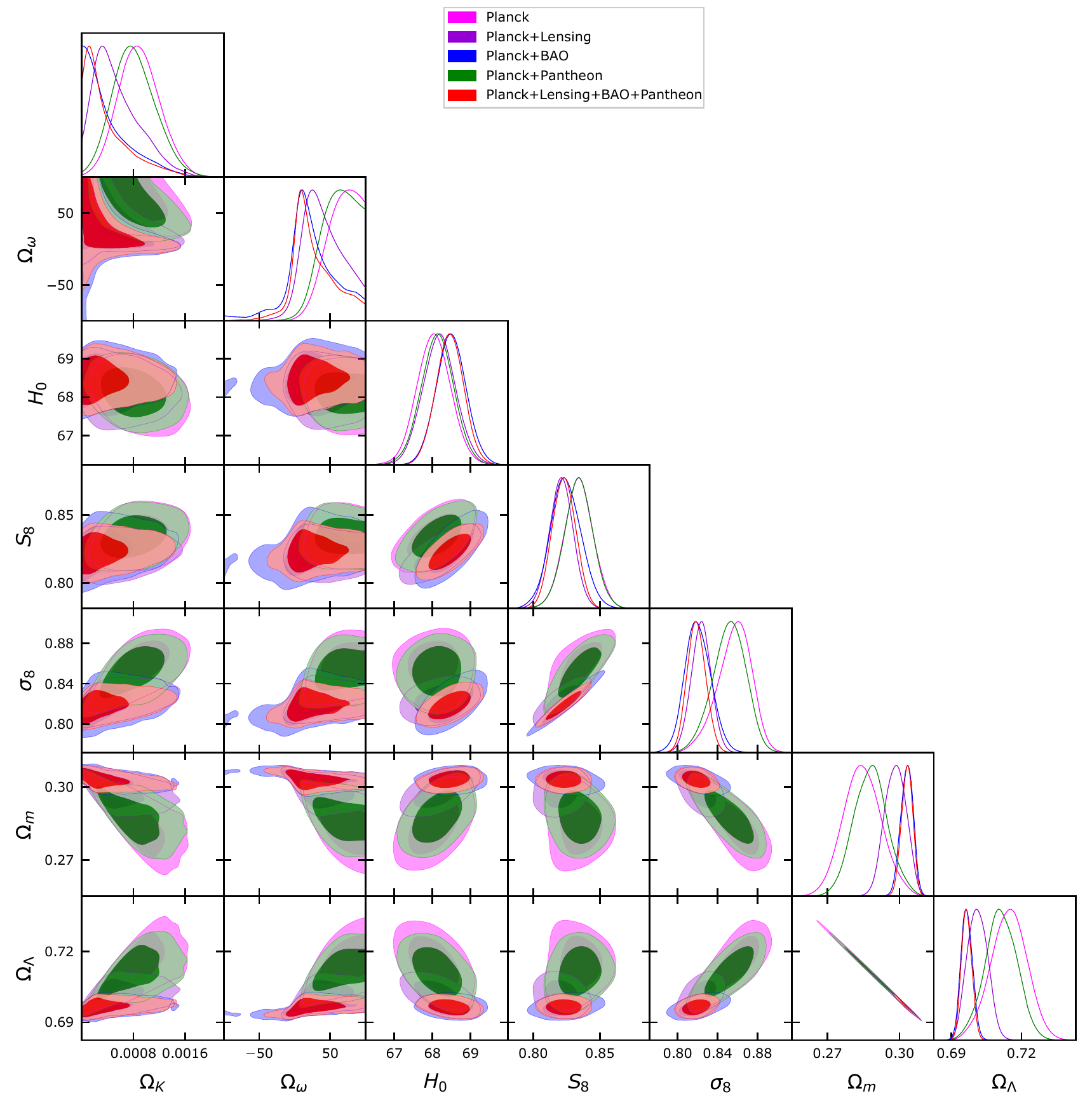}
\caption{
\label{fig:open}
Triangular plot showing 2D contours at 68\% and 95\% CL and 1D posterior distributions of a few key parameters in an open {HDE} universe. }
\end{figure*}


\begingroup
\squeezetable
\begin{center}
\begin{table*}[!]
\renewcommand{\arraystretch}{2}
\begin{tabular}{l|@{\hspace{0.5 cm}} ccccccccc}
\hline\hline
& Planck&Planck & Planck&Planck&Planck+Lensing \\
Parameters & &+Lensing & +BAO&+Pantheon&+BAO+Pantheon \\
\hline

$\Omega_b h^2$ & $    0.02256\pm0.00017$ & $    0.02248\pm0.00016$  & $    0.02239\pm0.00015$ & $    0.02253\pm0.00017$ & $    0.02240\pm0.00015$ \\

$\Omega_c h^2$ & $    0.1091\pm0.0047$ & $    0.1160\pm0.0033$ & $    0.1195\pm0.0021$ & $    0.1108\pm0.0042$ & $    0.1194\pm0.0021$  \\

$100\theta_{MC}$ & $    1.04132\pm0.00037$ & $    1.04111\pm0.00033$  & $    1.04101\pm0.00032$ & $    1.04127\pm0.00036$ & $    1.04100\pm0.00031$\\

$\tau$ & $    0.0520\pm0.0080$ & $    0.0519\pm0.0080$ & $    0.0549^{+0.0072}_{-0.0081}$ & $    0.0523\pm0.0079$ & $    0.0546\pm0.0075$ \\

${\rm{ln}}(10^{10} A_s)$ & $    3.032\pm0.017$ & $    3.034^{+0.018}_{-0.016}$  & $    3.044^{+0.015}_{-0.017}$ & $    3.034\pm0.017$ & $    3.043\pm0.015$ \\

$n_s$ & $    0.9729\pm0.0054$ & $    0.9695\pm0.0049$ & $    0.9721\pm0.0053$ & $    0.9716\pm0.0051$ & $    0.9669\pm0.0044$ \\

$\Omega_{k}$ & $    -0.00090^{+0.00034}_{-0.00029}$ & $    -0.00053^{+0.00046}_{-0.00016}$  & $    >-0.000504$ & $    -0.00083^{+0.00037}_{-0.00027}$ & $    >-0.000471$ \\

$\Omega_{\omega}$ & $    <-56.1$ & $    -41^{+34}_{-25}$  & $    -23\pm34$ & $    -62^{+16}_{-33}$ & $    -22\pm35$ \\

\hline

$\Omega_{\Lambda}$ & $  0.7150\pm0.0078$ & $  0.7011^{+0.0041}_{-0.0047}$  & $  0.6966^{+0.0023}_{-0.0026}$ & $  0.7118\pm0.0068$ & $  0.6964\pm0.0023$ \\

$\Omega_{m}$ & $    0.2859\pm0.0076$ & $    0.2994^{+0.0045}_{-0.0040}$  & $    0.3038^{+0.0025}_{-0.0022}$ & $    0.2891\pm0.0066$ & $    0.3040\pm0.0022$ \\

$\sigma_8$ & $    0.858^{+0.018}_{-0.014}$ & $    0.8232\pm0.0098$   & $    0.820^{+0.012}_{-0.014}$ & $0.852^{+0.018}_{-0.014}$ & $    0.8184\pm0.0094$ \\

$S_8$ & $    0.837\pm0.011$ & $    0.8223\pm0.0092$ & $    0.826\pm0.012$ & $    0.836\pm0.011$ & $    0.8239\pm0.0090$ \\

$H_0~[km/s/Mpc]$ & $   68.02\pm0.44$ & $   68.17\pm0.43$ & $   68.49\pm0.38$ & $   68.09\pm0.44$ & $   68.46\pm0.37$ \\

$r_{\rm{drag}}$ & $  147.29\pm0.31$ & $  147.31\pm0.31$  & $  147.13\pm0.30$ & $  147.27\pm0.31$ & $  147.15\pm0.29$ \\

\hline

$\chi^2_{\rm{bestfit}}$ & $  2755.6$ & $  2771.5$  & $  2770.5$ & $  3792.5$ & $  3815.4$ \\

$\Delta \chi^2_{\rm{bestfit}}$ & $ -10.4$ & $  -3.2$  & $ -1.4 $ & $  -8.6 $ & $ -1.6 $ \\


\hline\hline
\end{tabular}
\caption{Parameter constraints at 68\% CL for a {\it{closed}} universe on the independent (above the line) and dependent (below the line) parameters, together with the best-fit $\chi^2$ and its
difference from the flat LCDM.}
\label{tab:closed}
\end{table*}
\end{center}
\endgroup


\begin{figure*}
\centering
\includegraphics[width=\textwidth]{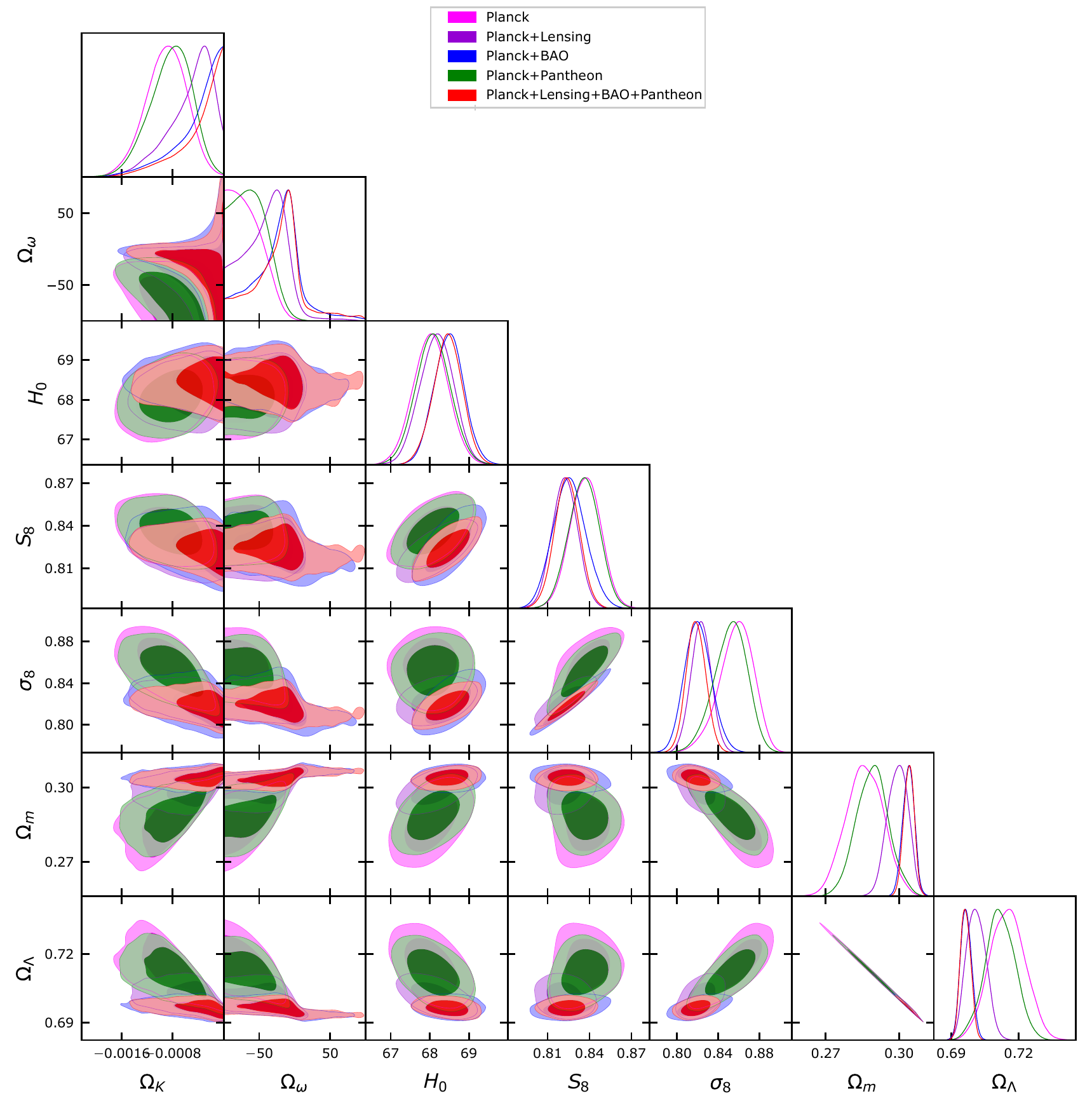}
\caption{
\label{fig:closed}
Triangular plot showing 2D contours at 68\% and 95\% CL and 1D posterior distributions of a few key parameters in a closed {HDE} universe.
}
\end{figure*}



\begingroup
\squeezetable
\begin{center}
\begin{table*}[!]
\renewcommand{\arraystretch}{2}
\begin{tabular}{l|@{\hspace{0.5 cm}} ccccccccc}
\hline\hline
& Planck&Planck & Planck&Planck&Planck+Lensing \\
Parameters & &+Lensing & +BAO&+Pantheon&+BAO+Pantheon \\
\hline

$\Omega_b h^2$ & $    0.02254\pm0.00017$ & $    0.02248\pm0.00016$  & $    0.02239\pm0.00015$ & $    0.02254\pm0.00017$ & $    0.02241\pm0.00015$ \\

$\Omega_c h^2$ & $    0.1092\pm0.0048$ & $    0.1157\pm0.0033$ & $    0.1193\pm0.0022$ & $    0.1104\pm0.0043$ & $    0.1191\pm0.0022$  \\

$100\theta_{MC}$ & $    1.04130\pm0.00036$ & $    1.04112\pm0.00033$  & $    1.04100\pm0.00032$ & $    1.04125\pm0.00035$ & $    1.04100\pm0.00031$\\

$\tau$ & $    0.0522\pm0.0077$ & $    0.0517\pm0.0080$ & $    0.0552\pm0.0080$ & $    0.0530\pm0.0079$ & $    0.0548\pm0.0074$ \\

${\rm{ln}}(10^{10} A_s)$ & $    3.033\pm0.018$ & $    3.034\pm0.017$  & $    3.044\pm0.017$ & $    3.035\pm0.017$ & $    3.043\pm0.015$ \\

$n_s$ & $    0.9727\pm0.0053$ & $    0.9698\pm0.0050$ & $    0.9668\pm0.0046$ & $    0.9723\pm0.0051$ & $    0.9670\pm0.0044$ \\

$\Omega_{k}$ & $    0.00000\pm0.00095$ & $    -0.00001\pm0.00068$  & $    -0.00001\pm0.00054$ & $    -0.00002\pm0.00090$ & $    -0.00001\pm0.00053$ \\

$\Omega_{\omega}$ & $   {\rm  unconstrained}$ & $    -1\pm48$  & $    -2\pm43$ & $   {\rm  unconstrained}$ & $    -2\pm42$ \\

\hline
$\Omega_{\Lambda}$ & $  0.7143\pm0.0078$ & $  0.7013\pm0.0043$  & $  0.6966^{+0.0023}_{-0.0027}$ & $  0.7118\pm0.0068$ & $  0.6965^{+0.0020}_{-0.0023}$ \\

$\Omega_{m}$ & $    0.2857\pm0.0079$ & $    0.2987\pm0.0043$  & $    0.3034^{+0.0028}_{-0.0023}$ & $    0.2882\pm0.0068$ & $    0.3035^{+0.0024}_{-0.0022}$ \\

$\sigma_8$ & $    0.857^{+0.019}_{-0.015}$ & $    0.824\pm0.010$   & $    0.821^{+0.012}_{-0.014}$ & $0.852^{+0.018}_{-0.014}$ & $    0.8185\pm0.0092$ \\

$S_8$ & $    0.836\pm0.011$ & $    0.8218\pm0.0090$ & $    0.825\pm0.012$ & $    0.835\pm0.011$ & $    0.8232\pm0.0088$ \\

$H_0~[km/s/Mpc]$ & $   68.06\pm0.44$ & $   68.17\pm0.42$ & $   68.49\pm0.39$ & $   68.07\pm0.43$ & $   68.43\pm0.37$ \\

$r_{\rm{drag}}$ & $  147.28\pm0.31$ & $  147.31\pm0.31$  & $  147.12\pm0.30$ & $  147.28\pm0.30$ & $  147.17\pm0.29$ \\

\hline

$\chi^2_{\rm{bestfit}}$ & $  2755.6$ & $  2771.5$  & $ 2770.5 $ & $  3792.7 $ & $ 3815.4 $ \\

$\Delta \chi^2_{\rm{bestfit}}$ & $ -10.2$ & $  -3.1$  & $ -1.4 $ & $  -8.4 $ & $ -1.6 $ \\


\hline\hline
\end{tabular}
\caption{Parameter constraints at 68\% CL for a full $\Omega_k$ universe on the independent (above the line) and dependent (below the line) parameters, together with the best-fit $\chi^2$ and its difference from the flat LCDM.}
\label{tab:full}
\end{table*}
\end{center}
\endgroup


\begin{figure*}
\centering
\includegraphics[width=\textwidth]{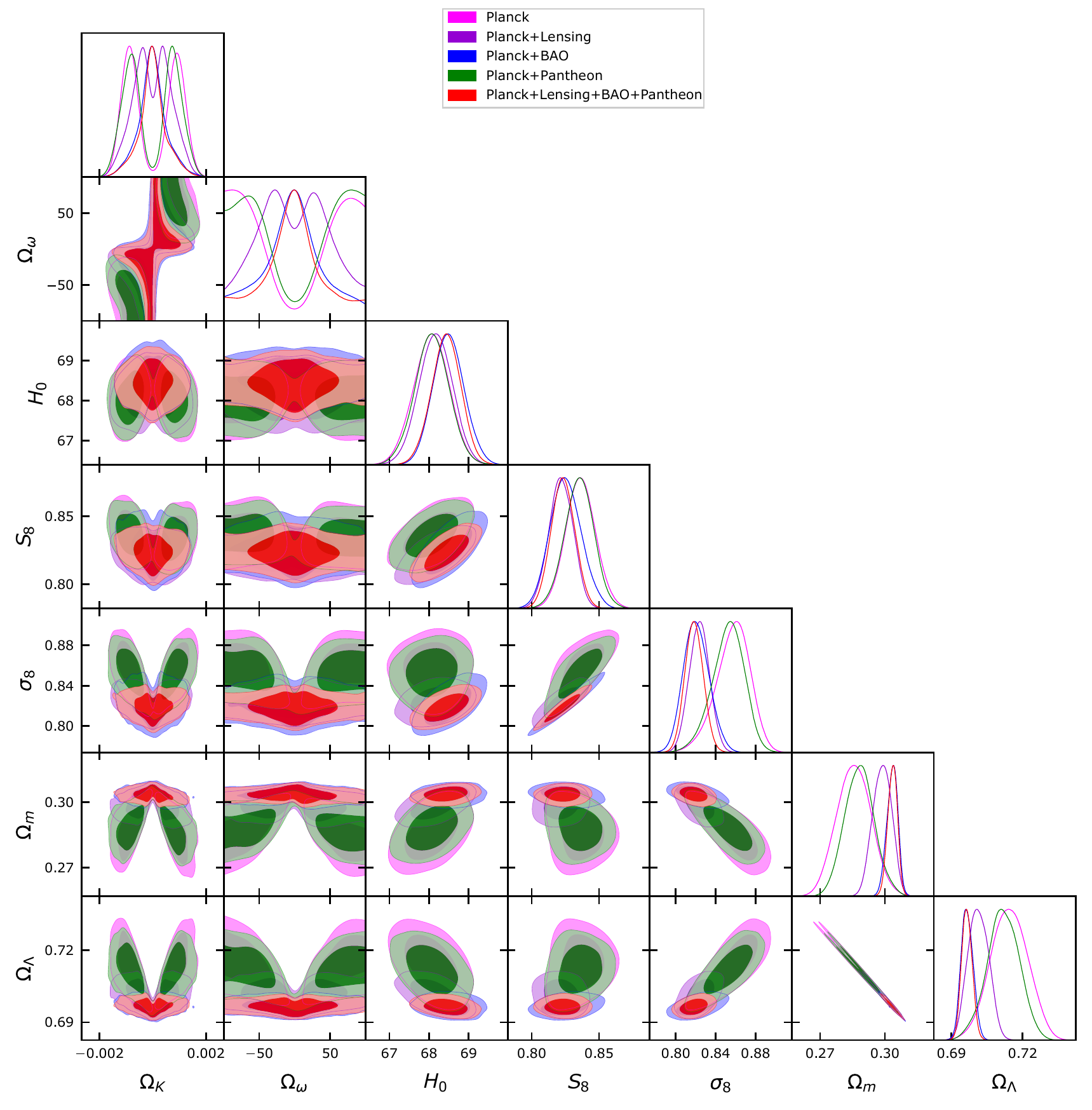}
\caption{
\label{fig:full}
Triangular plot showing 2D contours at 68\% and 95\% CL and 1D posterior distributions of a few key parameters in the full $\Omega_k$ universe.
}
\end{figure*}


\section{Results}
\label{sec:results} 

In this section, we present the results we obtained for the HDE model
in the three different cases: an open universe in Table~\ref{tab:open} and Fig.~\ref{fig:open}, a closed universe in Table~\ref{tab:closed} and Fig.~\ref{fig:closed}, and the full range for positive and negative values of the curvature in Table~\ref{tab:full} and Fig.~\ref{fig:full}. For completeness, 
we test in 
Appendix~\ref{cs2} what happens with a $\hat{c}_s^2$ free to vary
as an additional parameter.  Moreover, for comparison, we include similar Tables~\ref{tab:lcdm} and~\ref{tab:klcdm} with the same data-sets combination 
for the flat and non-flat LCDM models, respectively,
in Appendix~\ref{LCDM}. In each Table we present the constraints on the cosmological parameters at 68\% confidence level (CL), and in the last 2 rows, we report the best-fit, {\it i.e.}, minimum, $\chi^2$ and its difference from the flat LCDM model as described above.
Our noticeable results are as follows: \\

1. Regarding $\Omega_k$ in Tables~\ref{tab:open} and~\ref{tab:closed}, we see, for Planck alone or Planck+Pantheon, a preference for an open (closed) universe at more than 95\% CL (see also Figs.~\ref{fig:open} and~\ref{fig:closed}).
However, when we include the Lensing likelihood we see that the preference for an open (closed) universe is about just $1\sigma$. Moreover, when we add the BAO data, the lower bound of $\Omega_k$ is constrained to be close to zero and it can be consistent with a flat universe. We also note that the constraints on $\Omega_k$ are not Gaussian. The preference for a non-flat universe for Planck alone is similar to the standard non-flat LCDM (see Table~\ref{tab:klcdm} in Appendix~\ref{LCDM}). But, contrary to the preference of a {\it closed} universe in the LCDM scenario~\cite{Planck:2018vyg,DiVa:2019,Hand:2019,DiValentino:2020hov,Yang:2022kho,Semenaite:2022unt}, the two separate cases with positive and negative curvature are almost symmetric and moreover, as we can notice from the best-fit $\chi^2$, they are equally probable so that there is no preference of one case over the other. 
This will be the reason why the full case shown in Table~\ref{tab:full} gives for the $\Omega_k$
the average of the two separate cases, preferring therefore 
an almost null value $\Omega_k \approx 0${, which gives back the flat LCDM,} with an accumulated
uncertainty which is very small but enough to nullify the average value, though even better with BAO.\\

2. $\Om_\om$ is a newly introduced parameter in our model with
no {\it a priori} known constraints (cf.~\cite{Nils:2021}).
Our results show an intimate relation of $\Om_\om \propto \Omega_k$ and the properties for $\Omega_k$ still apply to $\Om_\om$ also, {\it e.g.} for Planck alone or Planck+Pantheon, a preference for $\Om_\om>0$ ($\Om_\om<0$) in an open (closed) universe, whilst $\Om_\om$ is more poorly constrained than $\Omega_k$. Actually, one can easily notice that 2D contours for the cut of $\Om_\om>0$ in Fig.~\ref{fig:open} (or $\Om_\om<0$ in Fig.~\ref{fig:closed}) can be mapped onto the corresponding 2D contours for $\Omega_k$, while the other contours for $\Om_\om<0$ (or $\Om_\om>0$) rapidly decays to zero\footnote{A similar behavior for $\Omega_k$ is expected if the positive and negative values of $\Omega_\om$ are considered separately.} and are believed to be numerical errors. This property will explain the similarities between 2D contours involving $\Omega_k$ and $\Omega_\om$ for the full $\Omega_k$ case shown in Fig.~\ref{fig:full}. Moreover, within our results alone, there is no preference of $\Om_\om<0$ over $\Om_\om>0$ or vice versa, just as for $\Omega_k$. This is the reason why we find $\Omega_\omega$ unconstrained for Planck and Planck+Pantheon data, while it is constrained to be close to zero for all the remaining dataset combinations. However, if we choose $\Omega_\om<0$ from astrophysical arguments -- the absence of a {\it complex} metric inside a black hole with a positive cosmological constant $\Lambda>0$~\cite{Argu:2015},\footnote{The required condition may be written as $\Omega_\om (\Omega_\om-2 \Omega_\Lambda)>0$~\cite{Argu:2015}. If we rule out the possibility of $\Omega_\om>0, \Omega_\om>2 \Omega_\Lambda$ in our case (see Fig.~\ref{fig:open}), we have only the choice of $\Omega_\om<0$ from $\Omega_\Lambda>0$.} the above relation $\Om_\om \propto \Omega_k$ makes us choose a closed universe, {\it i.e.} $\Omega_k<0$, consistently with our preferred result in the earlier work~\cite{Nils:2021}.
\\

3. Regarding the cosmic tensions involving 
the Hubble constant $H_0$ and cosmic shear parameter $S_8$, we obtain a positive result because we can break the correlation between them: we have a shift of $H_0$ towards a higher value by $1 \sigma$, though not enough to solve the Hubble constant tension, leaving the value of the cosmic shear $S_8$ unaltered (see for comparison the flat and non-flat LCDM cases in Table{s~\ref{tab:lcdm} and}~\ref{tab:klcdm}). This is in contrast to the exacerbated tension for a non-flat LCDM, where $H_0=54.4^{+3.3}_{-4.0}~{\rm km/s/Mpc}, S_8=0.981 \pm 0.049$ (see Table~\ref{tab:klcdm}) for Planck alone, with a decreasing $H_0$ but increasing $S_8$, as well as other models for improving $H_0$~\cite{Knox:2019rjx,Jedamzik:2020zmd,DiValentino:2021izs,Perivolaropoulos:2021jda}, because in the HDE case we do not see the noticeable correlation between $\Omega_k$ and $H_0$ (or $S_8$) (see Figs.~\ref{fig:open} and~\ref{fig:closed}) that is present in the non-flat LCDM case.
Moreover, our results for different data sets show that $S_8$ has a shift towards a lower value, in agreement with the non-flat LCDM case, when we add the BAO in the data-set combinations, while $H_0$ has a $1\sigma$ shift towards a higher value. However,
we can see the usual positive correlation between $H_0$ and $S_8$ in each data set, so that $S_8$ is increased as $H_0$ is increased. 
On the other hand, as we can notice in our results, Tables~\ref{tab:open},~\ref{tab:closed}, and~\ref{tab:full}, this behaviour
does not depend on the curvature. \\

\begin{figure*}
\centering
\includegraphics[width=0.4\textwidth]{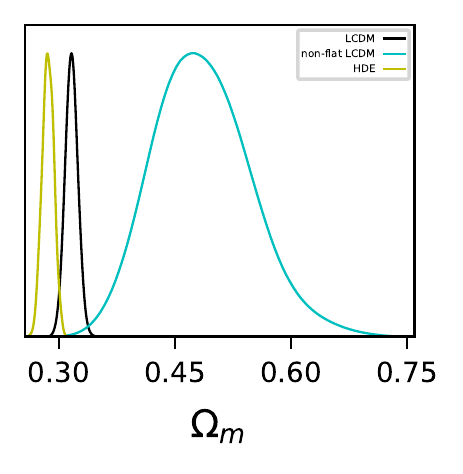}
\includegraphics[width=0.4\textwidth]{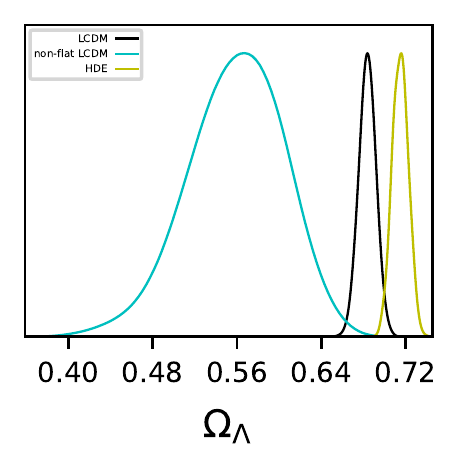}
\includegraphics[width=0.4\textwidth]{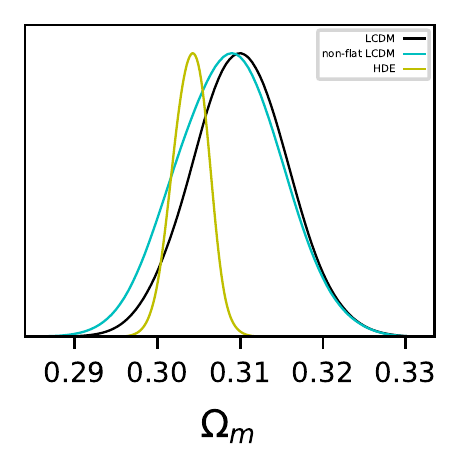}
\includegraphics[width=0.4\textwidth]{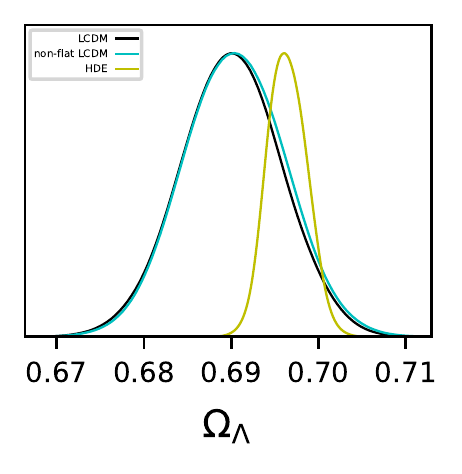}
\caption{
\label{fig:OL_vs_Om}
1D posteriors on $\Omega_m$ and $\Omega_\Lambda$ for
a closed universe scenario in our HDE model vs.
the flat and non-flat LCDM models for Planck (top panels), Planck+Lensing+BAO+Pantheon (bottom panels). 2D contour on $\Omega_m$ vs. $\Omega_\Lambda$ is {almost} a line 
from $\Omega_{\Lambda}\approx 1-\Omega_m$ (see Figs.~\ref{fig:open},~\ref{fig:closed}, and~\ref{fig:OL_vs_Om}).}
\end{figure*}

4. For all other parameters, there are some significant shifts, especially the matter density $\Omega_m$ and the dark energy density $\Omega_\Lambda$ with respect to a flat LCDM for all dataset combinations.
However, our results are more similar to the conventional value $\Omega_m \sim 0.3$ and $\Omega_\Lambda \sim 0.7$, contrary to the standard
non-flat LCDM result that shows $\Omega_m \sim 0.5$ and $\Omega_\Lambda \sim 0.5$ for Planck alone~\cite{Planck:2018vyg,DiVa:2019,Hand:2019,DiValentino:2020hov,Yang:2022kho}. 
For the other parameters, like $\Omega_b h^2$, $\Omega_c h^2$, $\theta_{MC}$, $\tau$, $A_s$, $n_s$, and $r_{\rm{drag}}$ that are not shown in Figs.~\ref{fig:open},~\ref{fig:closed}, and~\ref{fig:full}, there are a few shifts for Planck, Planck+Pantheon, or Planck+Lensing, but once BAO data is included and the parameter degeneracies are broken, the HDE values are very similar to the LCDM ones.
This may give a positive indication for our HDE model against a non-flat LCDM, even though there are no important improvements in the best-fit $\chi^2$ (see Table~\ref{tab:klcdm}), but there are against the LCDM case, as shown in the difference of the best fit $\chi^2$ from the flat LCDM (see Tables~\ref{tab:open}, \ref{tab:closed}, and \ref{tab:full}). An exception is the Planck+Pantheon case, where HDE performs significantly better than both LCDM and non-flat LCDM.
Hence, even if our model is assuming curvature in the universe, the results are close to {\it cosmic concordance}, {\it i.e.}, consistency with different cosmic observations (see Figs.~\ref{fig:open},~\ref{fig:closed}, and~\ref{fig:OL_vs_Om}) and they do not depend on the curvature.\\


\section{Concluding Remarks}

In conclusion, we have tested the {\it perturbed} dynamical dark energy model in Ho\v{r}ava gravity (HDE) due to an {\it effective} energy-momentum tensor from the extra Lorentz-violating terms. By treating the dark energy perturbations over the background perfect-fluid HDE as general fluid perturbations, we perform the full CMB data analysis via CAMB/CosmoMC as well as BAO and SNe Ia data. 
Except for the BAO case, we have obtained the preference for a {\it non-flat} universe, though the sign of the curvature parameter is not determined unless
we use additional arguments.
Thus, regarding the curvature parameter $\Om_k$, BAO is not consistent with other observations and this could
indicate some {\it flat} biases of BAO data points used in our analysis~\cite{Glanville:2022xes}.
On the other hand, we obtain some positive
results which seem to indicate that we are in the right direction toward a resolution of cosmic tensions. 
First, we obtained a positive result on the internal cosmic tension between the Hubble constant $H_0$ and cosmic shear parameter $S_8$ since we have a shift of $H_0$ towards a higher value by $1 \sigma$, though not enough for resolving $H_0$ tension, but the value of 
the cosmic shear $S_8$ is unaltered. 
This is in contrast to a decreasing $H_0$ but increasing $S_8$ in non-flat LCDM~\cite{Planck:2018vyg,DiVa:2019,Hand:2019,DiValentino:2020srs,DiValentino:2020hov}.
Second, for all other parameters, we obtain comparable results to those of LCDM 
especially with BAO, {\it e.g.}, $\Omega_m \sim 0.3$ and $\Omega_\Lambda \sim 0.7$, so that our results are close to a {\it cosmic concordance}, contrary to a recent non-flat LCDM result.
However, our results show also some undesirable features compared to our previous background analysis with the CMB distance priors~\cite{Nils:2021}, like (3) less improvement of the $H_0$ tension itself, (4) degeneracy between ($\bar{\om}, {\cal K}$) $\lra$ $(-\bar{\om}, -{\cal K})$, and the resulting almost {\it null} results on $\bar{\om}, {\cal K}$ or equivalently $\Om_{\om}, \Om_k$ if we do not restrict to ${\cal K}\neq 0$.
Several promising
directions for improving our analysis are:\\

1. The degeneracy between
($\bar{\om}, {\cal K}$) $\lra$ $(-\bar{\om}, -{\cal K})$
is the characteristic feature of $\Om_{\rm DE}$ for the \Ho~cosmology
background and there are some remnants in our previous background analysis
(model B) as well~\cite{Nils:2021}, though not quite as strong as in the current
analysis. However, in another HDE model (model A), which has a parameter
$\De N_{\rm eff}$ representing a possible excess in the standard effective
number of relativistic species $N_{\rm eff}=3.044$, the degeneracy is removed and
we have obtained (a) non-null results on $\Om_{\om}, \Om_k$ with the preference
of a closed universe and (b) a more improved $H_0$ tension. So, extending our
present analysis with $\De N_{\rm eff}$
as in the dark energy model A~\cite{Nils:2021}, and/or varying $N_{\rm eff}$ can be a promising way to improve our results.
\\

2. There is still no direct observational evidence for the interaction between dark matter and dark energy. However, it seems that a hypothetical dark energy model with the interaction, which is called the ``interacting dark energy (IDE) model'' may provide another appealing solution to $H_0$ and $S_8$ tensions~\cite{DiValentino:2019ffd,DiValentino:2020kpf}. Actually, in our \Ho~gravity set-up, the interaction could be natural if we can introduce CDM from the gravity sector also, as proposed in footnote {No. 2}. So, extending our analysis with the phenomenological interaction parameter for dark matter and energy, even without knowing the detailed mechanism, can be also an interesting way to improve our results.
\\

3. In the cosmological perturbation around the spatially-flat FLRW background, the leading scale-invariant spectrum for the scalar mode depends on the combination of UV parameters $\widetilde{\al}_4 \equiv \al_4 +2 \al_5/3+8 \al_6/3$, which vanishes for the parameters from the detailed balance condition (DBC) (5). So, we need to relax the DBC for UV parameters to obtain a scale-invariant {\it scalar} power spectrum and the result
would be still valid in a non-flat universe since the non-flatness just gives some sub-leading corrections to the flat cosmology perturbations. However, our result from CAMB/CosmoMC shows an almost scale-invariant {\it matter} power spectrum as usual and does not seem to depend much on the choice of UV parameters. This seems to be an evidence that we have lost some of the genuine UV effects in the dark energy perturbations from the coarse-graining in our fluid approach. The undesirable features in our result might be due to this problem also.

So, considering the
full perturbation analysis for a non-flat universe
with the corresponding modifications in the 
CAMB code would be a challenging way for the improvement. Phenomenologically, it seems that the general analysis may correspond to relaxing a vanishing anisotropic stress condition $\sigma=0$ since this condition would be due to some peculiar way of cancelation of arbitrary perturbations which would be anisotropic generally. Considering a non-vanishing anisotropic stress condition~\cite{Hu:1998kj} can be also an interesting way for the improvement, within the current fluid approach.\\

{4. In this paper, we considered $\la=1$ for simplicity of our analysis. If we consider an arbitrary $\la \neq 1$, as we noted in footnote No. 5, the fundamental constants defined in the Einstein equations (\ref{Einstein_eq}) and the Friedmann equations (\ref{FF1}), (\ref{FF2}) are different and we need to consider the effective speed of light, Newton's constant, and cosmological constant $c^2_{\rm{eff}}=(8(3 \la-1)^2)^{-1} {\ka^4 \mu^2 \La_W}$, $G_{\rm{eff}}=(16 \pi (3 \la-1))^{-1} {\ka^2  c^2_{\rm{eff}} }$, 
${\Lambda}_{\rm{eff}}=({3}/{2}) \Lambda_{W} c^2_{\rm{eff}}$, respectively.  
 All the coupling constants $\la, \ka, \mu$, and $\Lambda_W$ could flow under renormalization group (RG) so that the fundamental constants could flow also in the cosmic evolution. The fully consistent treatment of these evolving constants is beyond the scope of this paper but one might estimate the amount of RG flow from the existing cosmic tensions. For example, if we assume that $c_{\rm{eff}}=c, {\Lambda}_{\rm{eff}}={\Lambda}$ as the current values and do not RG run but only $\la$ can run, then one can find $G_{\rm{eff}}=2(3 \la-1)^{-1} G$, where $G=(32 \pi)^{-1} \ka^2  c^2_{\rm{eff}}$ is the Newton's constant in the Einstein equations (\ref{Einstein_eq}) even for arbitrary $\la$ and coincides with the Newton's constant in the Friedmann equations (\ref{FF1}), (\ref{FF2}) for $\la=1$~\cite{Dutta:2010jh,Frusciante:2015maa,Frusciante:2020gkx}. In this simple example, $G$ either run or does not run depending on the running behaviors of $\ka$ and $\mu$: (i) if $\ka$=fixed, $\mu^2/(3 \la-1)^2$=fixed, we have $G$=fixed, $G_{\rm{eff}}\sim (3 \la-1)^{-1}$, or (ii) if $\mu$=fixed, $\ka^4/(3 \la-1)^2$=fixed, we have  $G_{\rm{eff}}$=fixed, $G\sim (3 \la-1)^{-1}$; if we choose $G$ as the current value and do not run as in the case (i), $G_{\rm{eff}}$ shows the {\it asymptotically free} behavior at $\la=1/3$, which is thought to be a UV fixed point \cite{Barvinsky:2019rwn}, whereas if we choose $G_{\rm{eff}}$ as the current value and do not run as in the case (ii), $G$ shows the asymptotically free behavior.   
 If we consider the spatially flat case ${\cal K}=0$, and neglect the small cosmological constant term by considering the early universe, one can find that the Hubble parameter $H(t)_{\la \neq 1}$
 has an additional factor compared to the Hubble parameter for $\la=1$ as  $H(t)_{\la \neq 1}\approx 2 (3 \la-1)^{-1} H(t)_{\la=1}$. If we consider the RG flow of $\la_{\rm{IR}}=1\ra \la_{\rm{UV}}=0.9$, we can obtain about $10 \%$ increase of the Hubble parameter $H(t)_{\la \neq 1}$ compared to what is expected for $\la=1$, $H(t)_{\la=1}$ (see also~\cite{Nilsson:2019bxv}).
 In other words, the Hubble tension might be an indication of RG flow on $\la$. Moreover, we would expect that all the (fundamental) perturbation equations, like the evolution equations for matter density contrast, are also governed by the effective constant $G_{\rm{eff}}$ as in the background Friedmann equations so that $S_8$ could be also affected. It would be interesting to see the effect of $\la \neq 1$ in the cosmic tensions with the full data set analysis and observe the indication of RG flows in our cosmic evolution. }

\begin{acknowledgments}
%
EDV is supported by a Royal Society Dorothy Hodgkin Research Fellowship.
NAN and MIP were supported by Basic Science Research Program through the National Research Foundation of Korea (NRF) funded by the Ministry of Education, Science and Technology {{(2020R1A2C1010372) [NAN]}, (2020R1A2C1010372, 2020R1A6A1A03047877) [MIP]}.
This article is based upon work from COST Action CA21136 Addressing observational tensions in cosmology with systematics and fundamental physics (CosmoVerse) supported by COST (European Cooperation in Science and Technology).
We acknowledge IT Services at The University of Sheffield for the provision of services for High Performance Computing.
\end{acknowledgments}

\section*{Data Availability}
The data sets used in this work to constrain the models are public data available in their
respective references.


\appendix
\section{Testing the HDE model with a varying $\hat{c}_s^2$} 
\label{cs2}

In this Appendix,
we test the HDE model with a varying $\hat{c}_s^2$ in the range [-10,10] as an additional parameter and we report the results in Table~\ref{tab:cs2} and Figure~\ref{fig:cs2}. As we can see, this additional parameter is completely unconstrained and uncorrelated from the other parameters of the model, making the canonical choice of $\hat{c}_s^2=1$ in our main data analysis justifiable.


\begingroup
\squeezetable
\begin{center}
\begin{table*}[h!]
\renewcommand{\arraystretch}{2}
\begin{tabular}{l|@{\hspace{0.5 cm}} ccccccccc}
\hline\hline
 & Planck&Planck+Lensing \\
Parameters & &+BAO+Pantheon \\
\hline

$\Omega_b h^2$ & $    0.02255\pm0.00017$ & $    0.02239\pm0.00015$ \\

$\Omega_c h^2$ & $    0.1094\pm0.0046$ &  $    0.1195\pm0.0022$  \\

$100\theta_{MC}$ & $    1.04130^{+0.00034}_{-0.00039}$ & $    1.04099\pm0.00032$\\

$\tau$ & $    0.0519\pm0.0080$ & $    0.0548\pm0.0077$ \\

${\rm{ln}}(10^{10} A_s)$ & $    3.032\pm0.011$  & $    3.043\pm0.015$ \\

$n_s$ & $    0.9729\pm0.0055$  & $    0.9670\pm0.0044$ \\

$\Omega_{k}$ & $    -0.00090^{+0.00035}_{-0.00029}$  & $    >-0.00047$ \\

$\Omega_{\omega}$ & $    <-55.0$ & $    -27^{+33}_{-24}$ \\

$\hat{c}_s^2$ & $    {\rm unconstrained}$ & $    {\rm unconstrained}$ \\

\hline

$\Omega_{\Lambda}$ & $    0.7145\pm0.0075$  & $    0.6964\pm0.0022$ \\

$\Omega_{m}$ & $    0.2864\pm0.0074$  & $    0.3040\pm0.0022$ \\

$\sigma_8$ & $    0.857^{+0.018}_{-0.015}$  & $    0.8190\pm0.0092$ \\

$S_8$ & $    0.837\pm0.011$  & $    0.8244\pm0.0090$ \\

$H_0~[km/s/Mpc]$ & $   68.04\pm0.45$  & $   68.47\pm0.37$ \\

$r_{\rm{drag}}$ & $  147.28\pm0.31$  & $  147.15\pm0.29$ \\

\hline

$\chi^2_{\rm{bestfit}}$ & $  2755.4$ & $  3815.3$ \\


\hline\hline
\end{tabular}
\caption{Parameter constraints at 68\% CL for a closed universe, once $\hat{c}_s^2$ is free to vary in the range [-10,10], on the independent (above the line) and dependent (below the line) parameters, together with the best-fit $\chi^2$.
}
\label{tab:cs2}
\end{table*}
\end{center}
\endgroup



\begin{figure*}
\centering
\includegraphics[width=\textwidth]{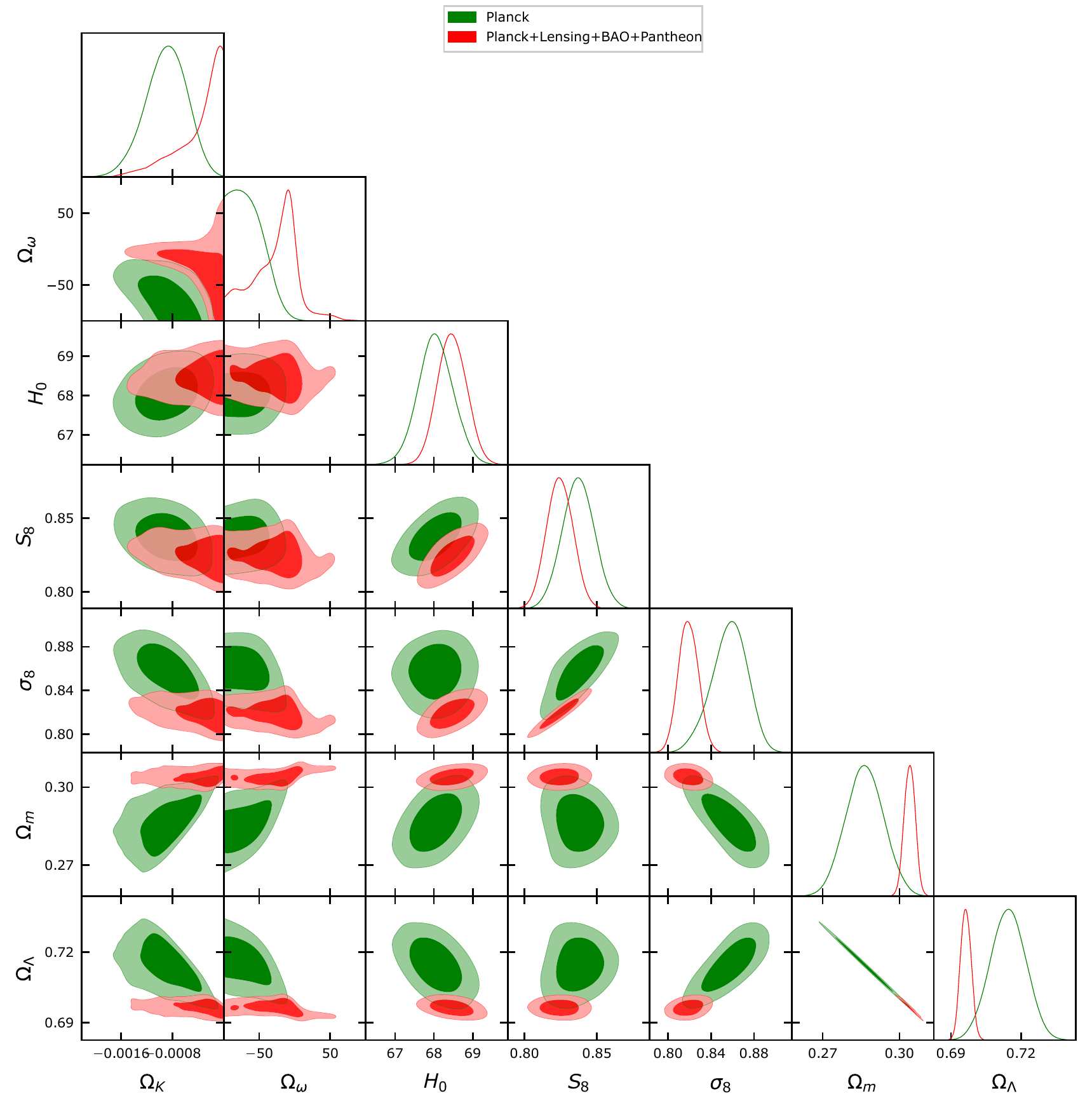}
\caption{
\label{fig:cs2}
Triangular plot showing 2D contours at 68\% and 95\% CL and 1D posterior distributions of a few key parameters {for a}
closed universe, once $\hat{c}_s^2$ is free to vary in the range [-10,10].
}
\end{figure*}


\begingroup
\squeezetable
\begin{center}
\begin{table*}[h!]
\renewcommand{\arraystretch}{2}
\begin{tabular}{l|@{\hspace{0.5 cm}} ccccccccc}
\hline\hline
& Planck&Planck & Planck&Planck&Planck+Lensing \\
Parameters & &+Lensing & +BAO&+Pantheon&+BAO+Pantheon \\
\hline

$\Omega_b h^2$ & $    0.02236\pm0.00015$ & $    0.02237\pm0.00015$  & $    0.02242\pm0.00014$ & $    0.02239\pm0.00014$ & $    0.02243\pm0.00013$ \\

$\Omega_c h^2$ & $    0.1202\pm0.0014$ & $    0.1200\pm0.0012$ & $    0.11933\pm0.00091$ & $    0.1199\pm0.0013$ & $    0.11921\pm0.00089$  \\

$100\theta_{MC}$ & $    1.04090\pm0.00031$ & $    1.04092\pm0.00031$  & $    1.04101\pm0.00029$ & $    1.04094\pm0.00031$ & $    1.04102\pm0.00029$\\

$\tau$ & $    0.0546\pm0.0078$ & $    0.0544\pm0.0073$ & $    0.0561\pm0.0071$ & $    0.0550\pm0.0078$ & $    0.0564\pm0.0071$ \\

${\rm{ln}}(10^{10} A_s)$ & $    3.045\pm0.016$ & $    3.044\pm0.014$  & $    3.047\pm0.014$ & $    3.046\pm0.016$ & $    3.047\pm0.014$ \\

$n_s$ & $    0.9648\pm0.0043$ & $    0.9649\pm0.0042$ & $    0.9665\pm0.0038$ & $    0.9655\pm0.0042$ & $    0.9668\pm0.0037$ \\

\hline

$\Omega_{\Lambda}$ & $  0.6834\pm0.0085$ & $  0.6847\pm0.0073$  & $  0.6889\pm0.0056$ & $  0.6855\pm0.0079$ & $  0.6897\pm0.0054$ \\

$\Omega_{m}$ & $    0.3166\pm0.0085$ & $    0.3153\pm0.0073$  & $    0.3111\pm0.0056$ & $    0.3145\pm0.0079$ & $    0.3103\pm0.0054$ \\

$\sigma_8$ & $    0.8122\pm0.0073$ & $    0.832\pm0.013$   & $    0.8102\pm0.0060$ & $0.8114\pm0.0074$ & $    0.8100\pm0.0060$ \\

$S_8$ & $    0.834\pm0.016$ & $    0.832\pm0.013$ & $    0.825\pm0.011$ & $    0.831\pm0.015$ & $    0.824\pm0.010$ \\

$H_0~[km/s/Mpc]$ & $   67.27\pm0.61$ & $   67.36\pm0.54$ & $   67.66\pm0.42$ & $   67.42\pm0.57$ & $   67.72\pm0.40$ \\

$r_{\rm{drag}}$ & $  147.05\pm0.30$ & $  147.09\pm0.26$  & $  147.21\pm0.23$ & $  147.11\pm0.29$ & $  147.23\pm0.23$ \\

\hline

$\chi^2_{\rm{bestfit}}$ & $  2765.8$ & $  2774.6$  & $  2771.9$ & $  3801.1$ & $  3817.0$ \\

\hline\hline
\end{tabular}
\caption{Parameter constraints at 68\% CL for a flat LCDM universe.}
\label{tab:lcdm}
\end{table*}
\end{center}
\endgroup

\begingroup
\squeezetable
\begin{center}
\begin{table*}[h!]
\renewcommand{\arraystretch}{2}
\begin{tabular}{l|@{\hspace{0.5 cm}} ccccccccc}
\hline\hline
& Planck&Planck & Planck&Planck&Planck+Lensing \\
Parameters & &+Lensing & +BAO&+Pantheon&+BAO+Pantheon \\
\hline

$\Omega_b h^2$ & $    0.02260\pm0.00017$ & $    0.02249 \pm 0.00016$  & $    0.02239 \pm 0.00015$ & $ 0.02247\pm0.00017   $ & $  0.02240\pm0.00015  $ \\

$\Omega_c h^2$ & $    0.1181 \pm 0.0015$ & $     0.1185 \pm 0.0015$ & $    0.1197 \pm 0.0014$ & $ 0.1190\pm0.0015  $ & $  0.1196\pm0.0014  $  \\

$100\theta_{MC}$ & $  1.04116 \pm 0.00033$ & $  1.04107 \pm 0.00032$  & $   1.04095 \pm 0.00031$ & $ 1.04102\pm0.00033 $ & $1.04096\pm0.00032 $\\

$\tau$ & $    0.0486 \pm 0.0082$ & $     0.0497^{+0.0082}_{-0.0071}$ & $    0.0548 \pm 0.0078$ & $ 0.0548\pm0.0077$ & $ 0.0559\pm0.0073 $ \\

${\rm{ln}}(10^{10} A_s)$ & $ 3.028 \pm 0.017$ & $ 3.030^{+0.017}_{-0.015}$  & $ 3.044 \pm 0.016$ & $3.043\pm0.016 $ & $3.047\pm0.014$ \\

$n_s$ & $     0.9706 \pm 0.0048$ & $     0.9688 \pm 0.0047$ & $     0.9659 \pm 0.0045$ & $0.9677\pm0.0047 $ & $0.9661^{+0.0042}_{-0.0047} $ \\

$\Omega_{k}$ & $  -0.044^{+0.018}_{-0.015}$ & $  -0.0106 \pm 0.0065$  & $   0.0008 \pm 0.0019$ & $-0.0064^{+0.0061}_{-0.0054} $ & $ 0.0008\pm0.0019 $ \\

\hline

$\Omega_{\Lambda}$ & $   0.560^{+0.050}_{-0.043} $ & $  0.659 \pm 0.017$  & $   0.6894 \pm 0.0061$ & $ 0.670\pm0.017 $ & $ 0.6901\pm0.0059$ \\

$\Omega_{m}$ & $     0.485^{+0.058}_{-0.068}$ & $    0.352 \pm 0.023$  & $     0.3098 \pm 0.0066$ & $ 0.337\pm0.022$ & $ 0.3087\pm0.0056$ \\

$\sigma_8$ & $   0.774 \pm 0.015$ & $    0.795 \pm 0.011$   & $     0.8109 \pm 0.0084$ & $0.8044\pm0.0096 $ & $ 0.8115\pm0.0072$ \\

$S_8$ & $     0.981 \pm 0.049$ & $    0.860 \pm 0.021$ & $    0.824 \pm 0.013$ & $0.852\pm0.025  $ & $ 0.824\pm0.010 $ \\

$H_0~[km/s/Mpc]$ & $    54.4^{+3.3}_{-4.0}$ & $    63.6^{+2.1}_{-2.3}$ & $    67.88 \pm 0.68$ & $ 65.1\pm2.2$ & $67.94\pm0.64 $ \\

$r_{\rm{drag}}$ & $   147.35 \pm 0.30$ & $  147.36 \pm 0.31$  & $  147.15\pm 0.31$ & $ 147.25\pm0.31$ & $ 147.16\pm0.29$ \\

\hline

$\chi^2_{\rm{bestfit}}$ & $  2754.5$ & $  2771.4$  & $  2771.4$ & $  3799.3$ & $ 3816.9 $ \\

$\Delta\chi^2_{\rm{bestfit}}$ & $  -11.3$ & $-3.2 $  & $ -0.5$ & $  -1.8$ & $ -0.1 $ \\
\hline\hline
\end{tabular}
\caption{Parameter constraints at 68\% CL for a non-flat LCDM universe, together with the best-fit $\chi^2$ and its difference from the flat LCDM.
}
\label{tab:klcdm}
\end{table*}
\end{center}
\endgroup

\section{Comparison of the flat and non-flat
LCDM models} 
\label{LCDM}

In this Appendix, we report for comparison the constraints on the cosmological parameters for the flat and non-flat LCDM models in Tables~\ref{tab:lcdm} and~\ref{tab:klcdm} for the same data-set combinations explored in this work.




\bibliography{main}

\end{document}